\documentclass{elsarticle}
\usepackage[T1]{fontenc}
\usepackage[utf8]{inputenc}
\usepackage{numprint}
\usepackage{paralist}
\usepackage{graphicx}
\usepackage{url}
\usepackage{caption}
\usepackage{subcaption}
\usepackage{booktabs}
\usepackage{multirow}
\usepackage{tcolorbox}
\usepackage{arydshln}
\usepackage[breaklinks, hidelinks]{hyperref}
\usepackage{hyperref}
\usepackage{xcolor}
\usepackage{xspace}
\usepackage{tabularx}
\usepackage{amsmath}
\npthousandsep{,}
\usepackage{rotating}
\usepackage{lineno,hyperref}
\usepackage{listings}
\modulolinenumbers[5]

\newcolumntype{b}{X}
\newcolumntype{s}{>{\hsize=.5\hsize}X}

\journal{Information and Software Technology}

\let\today\relax
\makeatletter
\def\ps@pprintTitle{%
    \let\@oddhead\@empty
    \let\@evenhead\@empty
    \def\@oddfoot{\footnotesize\itshape
         {Information and Software Technology. DOI: \url{https://doi.org/10.1016/j.infsof.2022.107082}} \hfill\today}%
    \let\@evenfoot\@oddfoot
    }
\makeatother

\newcommand{\approach}{{{MigrationExp\xspace}}}
\newcommand{\approachJtK}{{{MigrationExpJ2K\xspace}}}

\newcommand{\googlestrategy}{{{Google's guideline\xspace}}}

\newcommand{\rqtwoCapVI}{To what extent a ranking model learned from migrations done in real projects precisely recommends files to be migrated?}

\begin{document}
\begin{frontmatter}

\title{
Learning migration models for supporting incremental {language} migrations of software applications
}

%%Learning to migrate application

%\maketitle

\author{Bruno Góis Mateus}

\author{Matias Martinez\corref{cor1}}
\cortext[cor1]{Corresponding author}
\ead{matias.martinez@uphf.fr}

\author{Christophe Kolski}

\address{Université Polytechnique Hauts-de-France, France}
\address{LAMIH UMR CNRS 8201, France}
%\address[2]{Address of Smith K.}

%\author{Bruno Góis Mateus, Christophe Kolski, Matias Martinez\fnref{mycorrespondingauthor}}

%\author[mymainaddress,mysecondaryaddress]{Université Polytechnique Hauts-de-France}



\begin{abstract}

%\todo{To update}

{\bf{Context}:} 
A Legacy system can be defined as {a system that significantly resists modification and evolution}.
According to the literature, there are two main strategies to migrate a legacy system:
\begin{inparaenum}[a)]
\item to replace the legacy system by a new one, 
\item to incrementally migrate parts from the legacy system to the new one.
\end{inparaenum}
Incremental migration allows developers to better control the risks that may occur during the migration process.
However, this strategy is more complex because it requires decomposition of the legacy system into different parts, e.g. a set of files, and to define the order of migration of them along the migration process. To our knowledge, there is no approach to support developers on those activities.

{\bf{Objective}:} 
This paper presents an approach, named \approach{}, to support incremental {language} migrations of applications from one \emph{source} language to another \emph{target} language. 
\approach{} recommends the files that should be migrated first in a particular migration iteration.
As a novelty, our approach relies on a ranking model learned, using a \emph{learning-to-rank} algorithm, from migrations made by developers.

{\bf{Method}:} 
We validate our approach in the context of the migrations of Android apps, from Java to Kotlin, a new official language for Android.
We train our model using migrations of Java code to Kotlin written by developers on open-source applications.

{\bf{Results:}} 
%The results showed that, {on the task of proposing files to migrate}, our approach based on a ranking model learned from real migrations outperforms a migration strategy proposed by Google, which defines the order of migration based on files from an Android app.
{The results show that, on the task of proposing files to migrate, our approach outperforms a previous migration strategy proposed by Google, in terms of its ability to accurately predict empirically observed migration orders.}

{\bf{Conclusion:}} 
Since most Android applications are written in Java, we conclude that approaches to support developers such as \approach{} may significantly impact the development of Android applications.

\end{abstract}

\end{frontmatter}

%\linenumbers

\section{Introduction}

A legacy system can be defined as a system that is significantly resistant to modification and evolution~\cite{Brodie1995Legacy}.
Bisbal et al.~\cite{Bisbal1997Migration} mention that legacy systems can host problems because they usually run on obsolete hardware and lack clean interfaces to interact with other systems.
\begin{inparaenum}[\it 1)]
\item usually run on obsolete hardware,
\item maintenance can be expensive,
\item lack of clean interfaces to interact with other systems. %, and 
\item are difficult to extend.
\end{inparaenum}
Migration of such legacy systems offers more flexibility, better understanding of the system, easier maintenance, and reduced costs~\cite{Bisbal1997Migration}.

Brodie and Stonebraker discuss two main strategies to migrate a legacy system \cite{Brodie1995Legacy}. 
The first involves rewriting a legacy system from scratch to produce a new system (i.e., \emph{target} system) using modern software techniques and hardware of the target environment. We call it \emph{one-step migration}. Here, the legacy system remains operable until it is completely replaced by the target system.

The second strategy \emph{iterative} and \emph{incremental} migrates a system, in place, by small incremental steps until the desired long-term objective is reached. 
The iterative strategy involves incrementally selecting and migrating parts of the legacy system to become new parts of the incrementally constructed target system. 
During migration, the legacy system and the target system form a composite system that collectively provides all the functionalities \cite{Brodie1995Darwin}. 
Note that those migration approaches can not only be applied in legacy systems but in modernization of systems, for example, on the migration of applications initially written in Java and migrated to a modern Java-virtual machine language such as Scala, Kotlin, or Groovy. 

The incremental migration strategy has some advantages over a one-step migration~\cite{Brodie1995Legacy}.
The risk in the incremental is controllable, as it permits developers to control risk, step by step, by choosing the increment size: the smaller the increment, the smaller the risk.
If a step fails, only the failed step must be repeated, not the entire project, as in one step.
Moreover, one-step migration requires vast resources to completely rewrite an application from scratch, while in incremental migration, the required resources depend on the effort needed to execute one step.

However, a challenge in incremental migration is the decomposition of the legacy system into different parts, each of them are independent of the other ones and is migrated in a different migration step~\cite{Brodie1995Darwin}.
Brodie and Stonebraker \cite{Brodie1995Legacy} define different migration strategies, which consist of first decomposing the legacy system structure into different parts, each of those to be incrementally migrated. 
In particular, those strategies define steps from incrementally migrating 
\begin{inparaenum}[\it a)]
\item interfaces, 
\item applications and 
\item databases.
\end{inparaenum}

In this paper we focus on the iterative migration of \emph{applications}, which consists of decomposing the source code of an application and deciding the parts of the code to be migrated in a particular step.
In particular, we focus on \emph{language migration}, which consists of the migration of a piece of code written in a language to another language.
Previous strategies, including those of Brodie and Stonebraker~\cite{Brodie1995Legacy}, define steps or rules to guide developers during the migration process of an application. 
For example, Google provides a high-level guide to migrate Android applications from Java to Kotlin~\cite{AndroidDevelopers2020a}.
The selection of independent increments to migrate is one of the main challenges of iterative migration~\cite{Brodie1995Darwin}.

Unfortunately, those guidelines and strategies are too high-level because they define a migration plan of coarse-grained parts of the system under migration. 
For example, on Android migration defined by Google~\cite{AndroidDevelopers2020a} the migration order is given by the type of code entities: model class, tests, and utility functions.
For that reason, those fail to support developers on a fine-grained decomposition of the system, and thus a migration order of the decomposed parts. 
For instance, the Google guide mentioned does not include any strategy to define the migration order of model classes.

% https://medium.com/androiddevelopers/lessons-learned-while-converting-to-kotlin-with-android-studio-f0a3cb41669

% https://vaadin.com/blog/migrating-java-enterprise-apps-to-kotlin

% https://vaadin.com/blog/migrating-java-enterprise-apps-to-kotlin

% https://blog.ninja-squad.com/2018/05/22/kotlin-migration/

%https://medium.com/androiddevelopers/lessons-learned-while-converting-to-kotlin-with-android-studio-f0a3cb41669

The goal of this paper is to propose a novel study on the feasibility of building a migration model capable of supporting developers during incremental migration of an application. 
The goal of such a migration model is to suggest to developers the parts of the systems (e.g., files) to be migrated in a particular migration step. 
Our intuition is that it is possible to automatically build such a migration model from real migrations carried out by developers in the past.
The learned model captures how developers have migrated applications from a specific domain.
In our words, our goal is to lean a model that mimics the migration activity performed by developers.
To the best of our knowledge, no previous work has proposed a migration model of applications learned from migrations made by developers.

In this paper, we define a novel approach named \approach{} to support incremental file-level migrations based on a model learned from previous migrations.
Our approach is based on a \emph{ranking model}, which is trained with the goal of ranking, in a given migration iteration, the files to be migrated in that iteration: the files ranked at the top are those suggested to be migrated.

To validate the approach, we define an instance of our approach, named \approachJtK{}, to support the migration of Android applications from Java to Kotlin.
Kotlin is a multi-paradigm programming language, fully interoperable with Java, and adopted by Google as the official programming language for Android.
To train \approachJtK{}, we apply the supervised machine learning technique \emph{learning-to-rank}. %to create a model that suggests the files to migrate.
Our model is learned from real migrations from Java to Kotlin written by developers in \numprint{1457} open-source projects.

%The results of this paper show that, {on the task of proposing files to migrate}, \approach{} obtains a mean average precision (MAP metric~\cite{baeza1999modern}) higher than a strategy that follows Google's guideline~\cite{AndroidDevelopers2020a}.
{The results of this paper show that, on the task of proposing files to migrate, MigrationExp obtains a mean average precision (MAP metric [5]) higher than a strategy that follows Google's guideline [4], when using a ground truth of empirically observed migration orders.}
We consider that the proposed approach is an initial step towards a fully automated recommendation system to support applications' migration.

Our approach can be trained using different training datasets, instead of using migrations from Java to Kotlin extracted from open-source applications. 
For example, given a definition of good migration (e.g., causing less compilation and/or execution errors during the migration process), one could create a set of training samples that comply with such definition.

The contributions of this paper are as follows.
\begin{itemize}
    \item An approach that recommends migrations at the file level from one programming language to another.
    \item The materialization of that approach in the context of Java to Kotlin migration
    \item A benchmark of projects that performed migrations from Java to Kotlin.
   
%    \item A static analyzer tool that identifies 12 metrics exclusive to Android applications.\footnote{https://anonymous.4open.science/r/fe5cf980-060b-49ad-81b5-28de22f26360/}\todo{Is this important?? }
\end{itemize}

The paper continues as follows.
Section \ref{sec:term} explains the terminology used in the paper.
Section \ref{sec:motivation_example} presents a motivating example.
Section \ref{sec:context:kotlin} gives an overview of the Kotlin programming language in the context of Android development.
Section \ref{sec:approach} describes our approach and the instantiation of our approach in the context of Java to Kotlin migration.
Section \ref{sec:methodology} outlines the methodology used to evaluate our approach.
Section \ref{sec:evaluation} reports the evaluation results.
Section \ref{sec:threatsvalidity} presents the threats to validity.
Section \ref{cap6:sec:discussion} discusses the consequences of our results and future work.
Section \ref{sec:relatedwork} presents the related work.
Section \ref{sec:conclusion} concludes the paper.
All data presented in this paper is publicly available in our appendix: 
\url{https://github.com/UPHF/MigrationEXP}.

\section{Terminology}
\label{sec:term}

In this section, we present the terminology that we use in this paper in the context of programming language migration.

{\bf{Migration:}} the process of translating software from its \emph{source} programming language to the \emph{target} programming language.

{\bf{Migration step:}} A set of translations of code written in \emph{source} language to \emph{target} language that generates a \emph{new version} of an {operational} software.

{\bf{One-step migration:}} a migration process that \emph{fully} migrates a software application in one migration step
by rewriting a legacy system from scratch to produce the target system.
Therefore, applying this strategy, there is no reuse of any component of the legacy system~\cite{bisbal1997survey}.
In one-step migration, there is no version of the system written in the source and target programming language.
This kind of migration is known as the Big Bang~\cite{bateman1994migration} or the Cold Turkey~\cite{Brodie1995Darwin} migration.

{\bf{Incremental migration:}} a migration process that has more than one migration step.
It involves incrementally selecting and migrating parts of the legacy system to become new parts of the incrementally constructed target system\cite{Brodie1995Darwin}.
During this process, the versions of the operational software may have code written in both the source and target {programming} language.
This kind of migration is known as Chicken Little~\cite{Brodie1995Darwin}.

%%%%%%%%%%%%%%%
%by small incremental steps until the desired long term objective is reached. Each step requires a relatively small resource allocation (e.g., a few person years), a short time, and produces a specific, small result towards the desired goal. This is in sharp contrast to the vast resource requirements of a complete rewrite (e.g., hundreds of person years), a multi-year development, and one massive result. If a Chicken Little step fails, only the failed step must be repeated rather than the entire project. Since steps are designed to be relatively inexpensive, such incremental steps do not need to promise dramatic new function to get funded\cite{Brodie1995Darwin}

%A Chicken Little legacy IS migration involves incrementally selecting and migrating parts of the legacy IS to become new parts of the incrementally constructed target IS\cite{Brodie1995Darwin}

{\bf{Gateways:}} {during an incremental migration, the legacy and target systems interoperate to form the operational information system. 
This interoperability is provided by a module known, in general, as a gateway~\cite{Wu1997ButterflylegacyMigration}, a software module introduced between operational software components to mediate between them~\cite{Brodie1995Darwin}.}
{Gateways can play several roles in migration, insulating certain components from changes being made to others, translating requests and data between components or co-ordinating queries and updates between components~\cite{bisbal1997survey}.}
In some types of migrations, for example, migration of programming languages that are executed on the same platforms (e.g., Java virtual machine -JVM-), gateways are not required, because the language to be migrated (e.g., Java) can directly interact with code written in other JVM languages (e.g., Scala or Kotlin) and vice-versa.

{\bf{Language interoperability:}} the ability of two or more software components to cooperate despite differences in language, interface, and execution platform~\cite{Wegner1996}.

{\bf{File migration:}} the process of migrating a file from the \emph{source} language to the \emph{target} language.

{\bf{Commit with file migration:}} a commit that has one or more \emph{file migrations}.

\section{Motivating example}
\label{sec:motivation_example}
%https://vaadin.com/blog/migrating-java-enterprise-apps-to-kotlin
%As motivationa example, %%

{
The developer named Mahdy wrote an article\footnote{\url{https://vaadin.com/blog/migrating-java-enterprise-apps-to-kotlin}} on the blog of Vaadin framework (a web app development platform for Java\footnote{\url{https://vaadin.com/}}) which discusses the migration of an Android application from Java language to Kotlin, a new programming language promoted by Google for Android development (we focus on it in Section \ref{sec:context:kotlin}).
He chose to migrate \texttt{Beverage Buddy App}, a demonstration application written in Java that uses the Vaadin framework. The app code is publicly available on the Vaadin website\footnote{Vaadin example: \url{https://github.com/vaadin/beverage-starter-flow}}.
The app has 4 main packages:
\begin{inparaenum}[\it 1)]
\item \texttt{backend/},
\item \texttt{ui/common/},
\item \texttt{ui/views/}
and 
\item \texttt{ui/encoders}.
\end{inparaenum}
}

{
Mahdy, as other Android developers (e.g., \cite{duolingo_migration2020}), decided to perform an incremental migration (see the definition in Section~\ref{sec:term}), migrating Java files one by one using the autoconverter tools provided by IDEs such as Android Studio and IntelliJ IDEA. 
\emph{``What would be the next file to convert?''} Mahdy asked on the blog.
}

{
He started by choosing files from \texttt{ui/common/}, %such as the file \texttt{AbstractEditorDialog.java}.
but he found that not all Java files can be converted due to compilation issues related to, for example, \texttt{smart-casts}. %that required to be manually fixed.
Additionally, selecting Java classes from other packages such as \texttt{ui/views/reviewslist/} and converting them return other different issues, such as repeatable annotations and SAM (Single Abstract Method) conversion.
Fixing all those mentioned issues requires manual effort from developers.
Even worse, after fixing those issues, the project compiles, but fails to run properly, throwing the following exception:
}

\begin{minipage}{\linewidth}
\begin{lstlisting}[ basicstyle=\tiny]
Caused by: com.vaadin.flow.templatemodel.InvalidTemplateModelException: 
Element type '? extends org.vaadin.martin.backend.Review' is not a valid Bean type. 
Used in class 'ReviewsModel' with property named 'reviews' 
with list type 'java.util.List<? extends org.vaadin.martin.backend.Review>'.
    at com.vaadin.flow.templatemodel.BeanModelType.getListModelType(BeanModelType.java:271)
    at com.vaadin.flow.templatemodel.BeanModelType.getModelType(BeanModelType.java:177)
\end{lstlisting}
\end{minipage}

{
The developer found that this problem was related to the dependencies between classes:
Surprisingly, he found that if he converts one class (\texttt{Review.java}) before another one (\texttt{Category.java}) the app runs into code errors as that one shown, while converting in the opposite sequence compiles and runs fine.
}

{In this particular example, migrating  \texttt{Category.java} before \texttt{Review.java} is better because it allows developers to be more productive, as they do not have to spend time searching, applying and testing workarounds or fixes to avoid compilation or execution issues.}

{
As he wrote in the mentioned blog entry, those errors have been produced due to \emph{``the lack a perfect migration strategy''}.
This paper presents the first attempt to automatically produce an approach capable of helping developers choose the files to be migrated along the incremental migration of apps.
}

\section{Context: Kotlin and Migration of Android Application}
\label{sec:context:kotlin}

In this Section, we first briefly introduce the Kotlin programming language (Section \ref{sec:description_kotlin}).
Then, we present the relation between Android and Kotlin (Section \ref{sec:android_kotlin}) and about migration from Java to Kotlin (Section \ref{sec:context_migrating_kotlin})
Finally, we discuss the challenges of incremental migrations (Section \ref{sec:context_challenges_incremental}).

\subsection{What is Kotlin?}
\label{sec:description_kotlin}
%% Kotlin + Announced
In 2017, Google promoted Kotlin, a programming language that combines functional and object-oriented features, as an official Android language.
Kotlin is compiled to Java byte-code, which means that 
it is interoperable with Java, i.e., Kotlin code can invoke code written in Java and vice versa, both running on the same underlying JVM.

\subsection{Kotlin and Android}
\label{sec:android_kotlin}
In 2019, Google declared that Android became `Kotlin-first', which means that new APIs, libraries and documentation will target Kotlin and eventually Java~\cite{AndroidDevelopers2019}.
Since then, Google has advised developers to create new applications using Kotlin instead of Java~\cite{AndroidDevelopers2020b}.
However, thanks to the interoperability between Java and Kotlin, developers of Java-based Android applications do not need to migrate their apps fully to Kotlin, instead they can:
\begin{inparaenum}[i)]
\item add new Kotlin code and maintain the existing Java code, and/or
\item migrating some parts of their apps written in Java code to Kotlin.
\end{inparaenum}

This characteristic makes the adoption of Kotlin easier, and according to Google, in 2020 Kotlin was already used by over 60\% of professional Android developers, and 80\% of the top \numprint{1000} Android apps contain Kotlin code~\cite{AndroidDevelopersBetterApps}.

\begin{figure}[t]
    \centering
    \includegraphics[width=0.85\columnwidth]{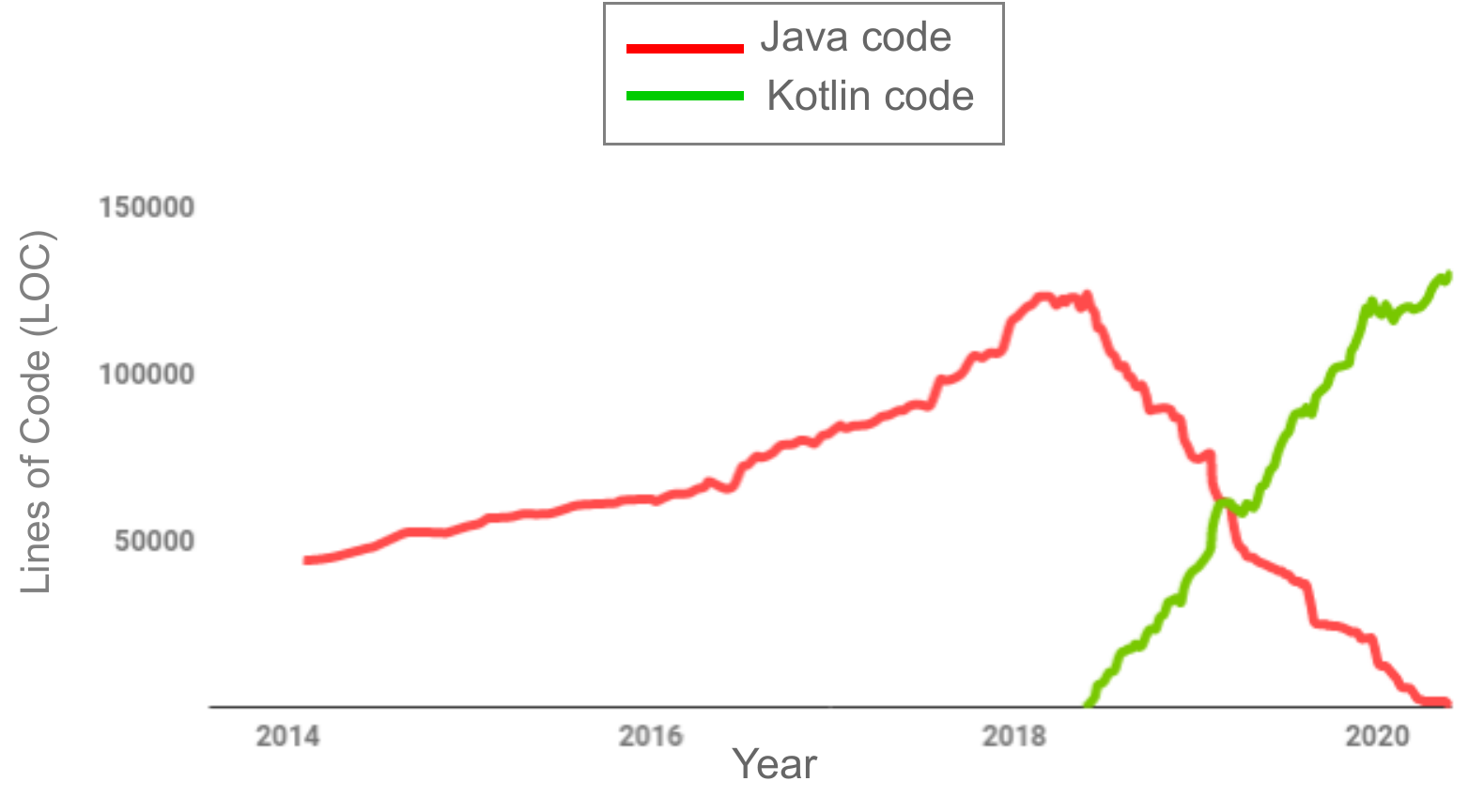}
    \caption{Evolution of the number of lines (LOC, axis X) of Java and Kotlin along with the Duolingo application's migration process~\cite{duolingo_migration2020} since 2014 (Axis Y).}
    %Axis Y shows the amount of code (SLOC) written, and axis X the time from 2014.}
    \label{cap6:fig:duo}
\end{figure}

\subsection{Migration of Java App to Kotlin}
\label{sec:context_migrating_kotlin}

The migration of Java to Kotlin has some peculiarities compared to legacy migration.
First, the underlying run-time environment (i.e., the Java virtual machine) or, in case of Android, the Android Runtime -ART- or Dalvik machines 
do not need to be updated: Kotlin and Java are both compiled to Java bytecode. 
Second, communication between migrated and non-migrated code in legacy migrations (e.g., COBOL to web~\cite{Colosimo2009Evaluatinglegacy}) needs \emph{wrappers} (\cite{Bisbal1997Migration,Colosimo2009Evaluatinglegacy}) or \emph{gateways} (\cite{Brodie1995Legacy, Bisbal1997Migration}).
The interoperability between Java and Kotlin means that wrappers and gateways are not necessary to migrate Java to Kotlin (Section \ref{sec:term}), especially focusing on incremental migration (Section \ref{sec:term}).

%\subsection{incremental migration of Android application to Kotlin}

%%%%
Moreover, some popular commercial Android applications also incrementally migrated from Java to Kotlin.
For example, Duolingo, a free science-based language education platform~\cite{duolingo_website}, was completely migrated in 2 years. 
Figure~\ref{cap6:fig:duo} shows the evolution of the amount of Java and Kotlin code from Duolingo.
During that period, Java files were progressively migrated to Kotlin, i.e., a commit migrated a subset of Java files, leaving other files in Java.

The \emph{incremental} migration allows developers  
to:
\begin{inparaenum}[\it a)]
\item migrate a subset of Java files, 
\item exhaustively test the migrated code to verify that the migrated code preserves the expected behavior,
and 
\item commit (and eventually release) a new version of their app before continuing with the migration of other files. 
\end{inparaenum}
As Duolingo's developers report~\cite{duolingo_migration2020}, incremental migration allowed them to apply strict testing, code review, and code style of each part of the application that was migrated.

%Recent research has shown that the 19\% of Android applications completely migrated from Java to Koltin were \emph{incrementally} migrated~\cite{GoisMateus2019}.

%Recent research has shown that the 19\% of Android applications completely migrated from Java to Koltin were \emph{incrementally} migrated~\cite{GoisMateus2019}.
%\todo{I dont have those numbers 19\%} \newc{ET 6 - Kotlin grows and Java decreases until the Java code is 0 is 19\%}

We have recently studied 374 open-source Android applications written, partially or totally, in Kotlin \cite{GoisMateus2019}.
We found that 86 applications were completely migrated from Java to Kotlin, and 55 of them (64\%) were incrementally migrated.
Moreover, we found that 214, initially written in Java, have been increasingly migrated, and the migrations have not been finished at the moment of writing this paper.

\subsection{Challenges of incremental migration}
\label{sec:context_challenges_incremental}

As reported by Brodie and Stonebraker~\cite{Brodie1995Legacy}, incremental migration faces several challenges.
This paper focuses on one of them:
given a version of the program to be migrated (composed of not yet migrated code and, eventually, some migrated code), a developer should select a set of files that she/he wants to migrate on that migration step.
This selection could be complex as:
\begin{inparaenum}[\it a)]
\item There could exist several candidate files to migrate, and 
\item the wrong selection of the files to be migrated could increase the migration effort due to emerging errors~\cite{AbdelAziz2020MigEnterprise,Nizet2018StoryMigration} or additional modifications to files not affected in the migration step. 
\end{inparaenum}
Unfortunately, existing migration guidelines provide high-level advice or guidance. For example, Google only suggests that migrations from Java to Kotlin on Android could start by migrating classes in the following order: class model first, then tests, utility functions, and, finally, other classes~\cite{AndroidDevelopers2020a}.
However, it does not include any guidance to help developers decide, for instance, which subset of model classes could be more convenient to migrate first. 

For these reasons, in the next section, we first present a migration recommendation system called~\approach{},  and its implementation called~\approachJtK{}, which focuses on the migration from Java to Kotlin.

\section{\approach: a recommendation system for supporting  incremental migrations
}
\label{sec:approach}

\subsection{Vision}
We envision a recommendation system that supports developers during incremental migration of the application $A$.
In each iteration $i$ of the migration of $A$, a portion of $A$ is migrated, while the rest is eventually migrated in future iterations.
Thus, the input of the system is $A$, conformed to parts that have not migrated and, eventually, already migrated parts.
The output of the system in $i$ is a list $L$ of \emph{parts} from $A$ that the system recommends to migrate in that iteration $i$.
The granularity of the recommendations, i.e., the part of the application that the approach recommends,  could vary according to the implementation: it could be a file, a package, a module, a subsystem, etc.

We envision such a system that 
\begin{inparaenum}[\it a)]
\item bases its decision exclusively on migrations previously done by developers from other migrated projects, and 
\item does not require any manually encoded migration rule.
This is a major difference from other recommendation systems based on encoded expertise~\cite{Jackson1986IntroductionTE}.

\end{inparaenum}

Given a set of migrations as training data, our system automatically learns a model used to recommend future migrations.
Each migration sample from the training data can be, for example, two versions of one application: one before the migration of one or more parts of the application, and the other the version which introduces the migration of those parts.

We materialize our vision in an approach named \approach{}, which focuses on programming language migration, i.e. projects that are migrated from a source language to a target language.
This section follows with a description of our approach.

\subsection{Architecture}

\begin{figure*}[ht]
    \centering
    \includegraphics[width=1\textwidth]{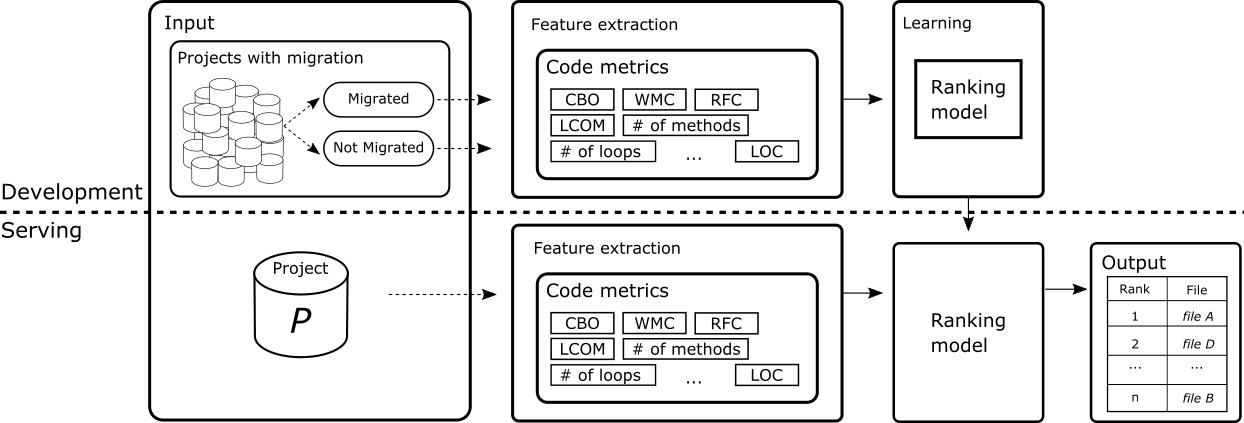}
    \caption{We apply two phases: \emph{development}, where we learn a ranking model, and \emph{serving} phase, where we use the model in production.}
    \label{fig:app:overview}
\end{figure*}

Our approach consists of two phases, as Figure~\ref{fig:app:overview} illustrates: 
\begin{inparaenum}[\it a)]
\item the development phase, and 
\item the serving phase. 
\end{inparaenum}
In the development phase, our approach learns a model from migrations from language $lang_1$ (e.g.,~Java) to $lang_2$ (e.g.,~Kotlin) written by, for instance, developers in open source projects.  
Then, in the serving phase, given a project $P$ as input, the model generated in the development phase is used to recommend file-level migrations: The model produces a list of candidate files to be migrated.
Now, we give a summary of both phases of our approach.

\subsubsection{Development phase}
\label{cap6:sec:design:approach:dev}

\approach{} relies on a learned model that encodes the migration performed by the developers.
In particular, we build a \emph{ranking model}, whose goal is to rank files not already migrated.
To train the model, we use a \emph{learning-to-rank} algorithm \cite{Liu2009}, which belongs to the family of supervised machine learning algorithms.
Our intuition is that we can build a ranking model that is able to capture the knowledge from developers to decide which file(s) migrates first given an app to be migrated.
A simplified illustrative example: if we train a model with projects in which developers have first migrated short files (expressed in SLOC), then our ranking model, given as input an app $Am$ to be incrementally migrated, will propose first to migrate the shortest files from $Am$. 

%we need to provide example data to train our model.
Each training point (an \emph{example}) used to train the model represents a migration done by a developer.
In our approach, each example is described by a \emph{vector} of measurements (or \emph{features}) and a label that denotes the category or class to which the example belongs~\cite{Hall1998}.

To create the training set used for training our ranking model, we extract data from projects that have been \emph{iteratively} migrated from one programming language to another (i.e., migrated in several commits).
In particular, for each commit $C$ of these projects, 
we first create a vector of features that represents each of those files.
The vector is composed of metrics extracted from those files.
Then, we put a \emph{label} on the vector: \emph{migrated} if the file represented by the vector was migrated by commit $C$, or \emph{not migrated}, otherwise.
These vectors are the training data used by our approach to learning a model.
Note that the labels are used by the learning-to-rank algorithm to learn the relation between migrated files and their features.

Finally, once the model is trained, it is then deployed and ready to be used in the serving phase.

\subsubsection{Serving phase}
\label{sec:serving}

In the serving phase, our approach \approach{} takes as input a program $P$, written partially or completely using $lang_1$, which the developers aim to migrate to $lang_2$. 
As done in the development phase, our approach extracts features from the project's files, i.e., candidate files to migrate, and creates for each file one vector of features, as done during the development phase.
These vectors are given as input to our model.
Finally, using this information, the model learned in the development phase ranks the project's files according to their relevance and returns the list of recommended files to be migrated.

%Figure~\ref{cap6:fig:ranking_overview} shows an example of this phase. 
Let us show a simple example to describe this phase.
The approach takes as input a project composed of 5 files where 4 files ($A.lang_1$, $B.lang_1$, $C.lang_1$ and $E.lang_1$) could be migrated, and one $D.lang_2$ already migrated.
The learned rank model ranks the four not yet migrated files based on the experience of developers by migrating similar files (i.e., similar vector features). 
In this example, the developer could start migrating the files at the top of the recommendation, e.g., $E.lang_1$, then testing the migrated app, committing the changes, and generating a new version to publish. 
Those activities form one step in the incremental migration process.
Note that in migrations that involve fully interoperable languages (e.g., Java and Kotlin) the migrated files (e.g., $D.lang_2$) could continue interacting with the not migrated (e.g., $B.lang_1$).
In case of working with non-interoperable languages, it would be necessary to build \emph{gateways} that interconnect the code written in those languages, as proposed by~\cite{Brodie1995Legacy}.

\subsection{\approachJtK{}: Supporting Migration from Java to Kotlin.}
\label{sec:approachJtK}

The approach \approach{} previously described is language independent.
In this section, we present an instantiation of the approach in the context of migrations of Java to Kotlin. 
This instance aims to help Android developers migrate from Java to Kotlin.

\subsubsection{Overview of \approachJtK{}}

 \approachJtK{} works at the level of files: it recommends Java files that can be manually or automatically (e.g., using a conversion tool provided by the IDE) migrated to Kotlin.

Given an application that should be iteratively migrated to Kotlin, our approach generates a rank with all candidate Java files to be migrated, where the top files are the recommendations to be migrated first, for instance, in the current migration iteration.
To create such an approach, we created a ranking model using a \emph{learning-to-rank} algorithm, which solves a ranking problem by sorting objects according to their degrees of relevance, preference, or importance~\cite{Liu2009}. 

In the remainder of this section, we first present how we use the information extracted from projects with file migration from Java to Kotlin to collect the data needed to build our ranking model (Section~\ref{cap6:sec:ranking:traning}).
Then, in Section~\ref{sec:representation}, we explain how we transform this data according to the representation used by learning-to-rank.
Finally, in Section~\ref{cap6:sec:design:features}, we describe the list of features extracted during the feature extraction process.

\subsubsection{Learning process for Java to Kotlin migration model}
\label{cap6:sec:ranking:traning}

%Our intuition is that we can build a learning-to-rank model that is able to capture from developers the knowledge to decide which file(s) migrates first given an app to be migrated.
%A simplified illustrative example: if we train a model with projects in which developers have migrated first short files (expressed in SLOC), then our ranking model, given as input an app $Am$ to be incrementally migrated, will propose first to migrate the shortest files from  $Am$. 

In this work, we automatically create a ranking model by feeding it with information from real migrations done by developers.
To this end, we use a learning-to-rank algorithm.
In learning-to-rank, the training data consists of queries and documents where each \emph{query} is associated with a set of \emph{documents}. 
The relevance of documents concerning the query is represented by a label~\cite{li2011short}.
In our context, each commit with at least one file migration from the training dataset becomes a \emph{query}.
A document associated with a query (and transitively to a commit $C$) corresponds to a file $f$, which belongs to the commit $C$.
Each query's documents are labeled with $1$ if the document (file) was migrated in the commit associated with the query. Otherwise, a document is labeled with $0$. % (when a file is not migrated by that commit).

To illustrate how we transform the information extracted from commits with migration in our training data set, imagine an application with three Java files ($File_1.java$, $File_2.java$, $File_3.java$).
Consider a commit that performs these actions: \begin{inparaenum}[\it i)]
\item removes ``$File_1.java$'',
\item updates ``$File_2.java$'', and
\item adds ``$File_1.kt$''.
\end{inparaenum}
This commit has a file migration ($File_1.java$ was migrated from Java to Kotlin).
Consequently, we label these documents as follows: $File_1.java$ as migrated (i.e., $1$), $File_2.java$, $File_3.java$ as not migrated (i.e., $0$).
From that information, we create a \emph{query}.

To prepare the data used to train the model, 
we create one query per commit that migrated code from our training set.
Finally, the set of queries is the input of the training process of the ranking model, which generates a learned ranking model as the output.

\subsubsection{Using Java to Kotlin migration model to support migration}
\label{cap6:sec:ranking:serving}

The learned ranking model is used in the serving phase (Section \ref{sec:serving}) to recommend migrations.
In that phase, the input is a query composed of files (documents) that belong to the application to be migrated. 
In fact, to obtain a recommendation, we create a query composed of those documents.
Note that those documents are not labeled.
Then, giving a query as input, the model outputs, for each document, a \emph{Predicted relevance} value.
By sorting these documents according to their values, from the most relevant to the least relevant, 
we obtain the ranking of recommendations, where the documents in the first positions are the ones to be prioritized during the migration.

\subsubsection{Representing documents and queries}
\label{sec:representation}

We now focus on the representation of files from a commit as documents belonging to a query.
Each file in a commit is represented by a \emph{vector} of features.
Consequently, a query is a set of vectors.
%The process of learning the model, which receives as input queries with labeled documents, learn the \emph{relation} between the features that represent the files and the labels (two in this paper: 1 for migrated and 0 for no migrated).
The process of learning the model receives as input queries with labeled documents, and has as goal to learn \emph{relations} between the features that represent the files and the corresponding labels (two in this paper: 1 for migrated and 0 for no migrated).

In the serving phase, we create a \emph{vector} for each file of the application to be migrated.
We create a query composed of a set of vectors, which is the input of the model. The model then ranks each vector (file) according to its label and features' value.

\subsubsection{Feature extraction for Java and Android apps}
\label{cap6:sec:design:features}

We extract features that represent and characterize the code that:
\begin{inparaenum}[\it a) ]
\item was already migrated, and
\item is under migration (i.e., not yet migrated).
\end{inparaenum}
The feature extraction phase receives as input a set of files, and we generate for each of them a vector of features. Then, those vectors are given as input to \approachJtK.
Each feature is created from a particular metric extracted from the source code files.
In total, we used 56 metrics that are listed in Table~\ref{tab:metrics}

First, we use 44 \emph{source code metrics} that have been defined and used in previous experiments related, for example, to the assessment of the overall quality of the software~(e.g., \cite{Eski2011,Peng2015,aniche2020effectiveness}). Table \ref{tab:metrics} shows them.
These metrics are grouped into different categories such as inheritance, communication, complexity and readability.
They include the object-oriented metrics proposed by Chidamber and Kemerer~\cite{Chidamber1994}, such as Weighted Methods per Class (WMC), 
readability metrics such as the number of loops and the number of comparisons proposed by Buse et al.~\cite{Buse2010} and Salabrino et al.~\cite{Scalabrino2017} and
other source code metrics such as the number of Sources Line Of Code (SLOC).

Second, we define 12 \emph{Android metrics} to capture the exclusive characteristics of the Android applications, which are presented in Table \ref{tab:Androidmetrics}

\begin{table}
\caption{List of Android metrics.}
\begin{tabularx}{\textwidth}{s b}
\toprule
\textbf{Metric} & \textbf{Description} \\
\midrule

\textit{isActivity}& a binary feature that informs whether a class extends the Activity class from the Android API.\\
\hline
 \textit{isView}& a binary feature that informs whether a class extends the View class from the Android API.\\
 \hline
 \textit{isBroadcastReceiver}& a binary feature that informs whether a class extends the BroadcastReceiver class from the Android API.\\
 \hline
 \textit{isService}& a binary feature that informs whether a class extends the Service class from the Android API.\\
 \hline
\textit{isContentProvider}& a binary feature that informs whether a class extends the ContentProvider class from the Android API.\\
\hline
\textit{isFragment}& a binary feature that informs whether a class extends the Fragment class from the Android API.\\
\hline
 \textit{isBuildingBlock}& a binary feature that informs whether a class extends one of the essential building blocks (Activity, Service, BrodcastReceiver and ContentProvider) of an Android application.\\
 \hline
\textit{isInAndroidHierarchy}& a binary feature that informs whether a class extends any class from the Android API.\\
\hline
\textit{Number of parameters coupled}& The number of methods parameters whose type is an object from the Android API.\\
\hline
\textit{Number of return coupled}& The number of methods whose return type is an object from the Android API.\\
\hline
 \textit{Number of methods coupled}& The number of methods whose at least one parameter or return type is an object from the Android API.\\
 \hline
\textit{hasAndroidCoupling}& a binary feature that informs whether a class has at least one method coupled.\\
\hline
\end{tabularx}
\label{tab:Androidmetrics}
\end{table}

%measurements are extracted from the data given as input to our approach to create vectors of features that compose our model's input.

We recall that the extracted features describe the source code under migration.
Then, from that data, we train a model that \emph{automatically}
\begin{inparaenum}[\it a)]
\item captures how the code under migration (represented by features) looks like, and 
\item learns the relation between migrated and non-migrated code through the extracted features.
\end{inparaenum}

%We may recall that the metrics to learn are not related to migration process but to describe the code that is under migration. For this reason, we decide to use metrics that represent code. Mention that we do not do previous feature engineering: No previous work has done that.

%To the best of our knowledge, no study establishes a relationship between metrics or measurements and source code file migrations. 
%\todo{TODO:We may recall that the metrics to learn are not related to migration process but to describe the code that is under migration. For this reason, we decide to use metrics that represent code. Mention that we do not do previous feature engineering: No previous work has done that.}
%For that reason, we decided to use 56 metrics as the features used by our approach to create a vector that represents a file from a project under migration. 
%These metrics are listed in Table~\ref{tab:metrics}.

\begin{table}
\caption{List of collected features grouped by category.}
\begin{tabular}{l l}
\toprule
\textbf{Category} & \textbf{Metric name} \\
\midrule
\multirow{2}{*}{Size} & Source Lines Of Code (SLOC),\\
& Number of methods, Number of fields \\\midrule
Complexity & Weight Method Class (WMC), Max nested blocks \\\midrule
\multirow{2}{*}{Coupling} & Coupling between objects (CBO),\\
& Response for a Class (RFC) \\\midrule
Encapsulation & Number of public fields, Number of public methods \\\midrule
\multirow{3}{*}{Cohesion} & Lack of Cohesion of Methods (LCOM),  \\
& Tight class cohesion (TCC),\\
& Loose Class Cohesion (LCC) \\\midrule
Inheritance & Depth Inheritance Tree (DIT) \\\midrule
\multirow{4}{*}{Readability}  & Number of unique words, Number of loops,\\
& Number of assignments,\\
& Number of comparisons, Number of string literals,\\
& Number of math operations, Quantity of numbers \\\midrule
\multirow{7}{*}{Android} & isActivity, isView, isBroadcastReceiver,\\
& isService, isContentProvider, isFragment,\\
& isBuildingBlock, isInAndroidHierarchy,\\
& hasAndroidCoupling, Number of methods coupled,\\
& Number of parameters coupled,\\
& Number of returns coupled \\\midrule
\multirow{14}{*}{Java-specific}  & Number of default fields, Number of default methods,\\
& Number of final fields, Number of final methods,\\
& Number of static fields, Number of static methods,\\
& Number of private fields, Number of private methods,\\
& Number of protected fields, {isPOJO}\\
& Number of protected methods,\\
& Number of abstract methods,\\
& Number of anonymous classes,\\
& Number of inner classes, Number of lambdas,\\
& Number of static invocation (NOSI), \\
& Number of synchronized fields,\\
& Number of synchronized methods,\\
& Number of parenthesized expressions,\\
& Number of returns, Number of try catches,\\
& Number of log statements, Number of variables \\\midrule
{Testing} & {isTest} \\\bottomrule
\end{tabular}
\label{tab:metrics}
\end{table}

\section{Methodology}
\label{sec:methodology}

This paper aims to evaluate the feasibility of using \approachJtK{} to help developers iteratively migrate Android applications.
The following research questions guide our study:

\begin{itemize}
\item  \emph{RQ: \rqtwoCapVI}
\end{itemize}

In this section, we present the methodology applied to answer this research question.
First, we present the method applied to collect open source applications that have performed migration of files from Java to Kotlin (Section~\ref{sec:met:data}).
Then, we describe how we learn a model from information about migrations performed by developers in these projects (Section~\ref{sec:met:training}).
Finally, in Section~\ref{cap6:sec:ranking:evaluation}, we explain how we evaluated the learned model.

\subsection{Data acquisition for training and evaluation} %%Data acquisition
\label{sec:met:data}

To train and evaluate \approachJtK, we need projects that have been migrated from Java to Kotlin.
For that, we created two datasets with Java to Kotlin migrations.
First, we collected migrations from an existing dataset of open-source applications written, partially or totally, in Kotlin, published on apps stores such as F-droid and Google Play.
With this dataset, namely $Android_{j2k}$,  we aim to train the model with the goal of capturing how Android developers migrate applications from Java to Kotlin.
Nevertheless, the model could be enriched with migrations from Java to Kotlin done in other types of projects (non-Android project).
For that reason, we collected additional migrated open-source applications hosted on GitHub. We call this dataset $GitHub_{j2k}$.

The usage of these two datasets allowed us to evaluate our model \emph{in the wild}, which is different from \emph{in the lab} (i.e., using one dataset to train and test our model applying 10-fold cross-validation), because it does not assume that 90\% of the domain knowledge is known beforehand~\cite{Allix2016}.

To build those datasets, we followed three steps:
\begin{inparaenum}[\it 1)]
\item identification of open source projects that use Kotlin, 
\item filtering projects that have Java code at any version, i.e., commits,  
and \item filtering projects that have migrated files from Java to Kotlin.
\end{inparaenum}

%We now detail how we built the two datasets of migrations.
%\paragraph{$GitHub_{j2k}$: dataset of open-source projects with migrations}

%We followed 3 steps to create our $GitHub_{j2k}$ dataset:

\textbf{Step 1. Identification of open source projects written in Kotlin.}
To build our dataset of Android applications with migrations, we extracted the FAMAZOA v3 repositories~\cite{FAMAZOA}.
FAMAZOA is a dataset of open source applications written in Kotlin, and it contains 387 applications written partially or totally in Kotlin collected from 3 datasets of open source Android applications: AndroidTimeMachine~\cite{Geiger2018:data}, AndroZoo~\cite{Allix2016} and F-Droid.\footnote{F-droid (repository of open-source Android applications):  \url{http://f-droid.org}}

Then, in order to create $GitHub_{j2k}$,  we searched on GitHub for repositories written in Kotlin.
Our search was performed on the publicly-available GitHub mirror available on Google BigQuery~\cite{hoffa_2016}.
This mirror offers a full snapshot of the content of more than 2.8 million open source repositories and almost 2 billion files. 
Moreover, it provides information about the use of programming languages in the last commit of each repository.
Therefore, we performed a query looking for projects that have Kotlin.
As a result, it returned \numprint{7119} repositories. 170 of them were already included in FAMAZOA, so we discarded them.

\textbf{Step 2. Identification of projects that used Java at its life-cycle.}
The previous step is necessary to identify projects that have Kotlin.
However, we needed to filter projects that contain Java as well, since this is a requirement to have migrations.
For that reason, we selected all projects with at least one commit with Java (i.e., a commit that introduces Java code).
At the end of this procedure, we identified \numprint{5126} repositories from GitHub and 270 from FAMAZOA.

%%We applied steps 2 and 3 presented in Section~\ref{sec:met:data}, and we identified 270 out of 387 (69\%) applications with at least one migration from Java to Kotlin.
%Since FAMAZOA includes applications hosted on GitHub, to avoid duplicates, we removed 170 applications from $GitHub_{j2k}$ that are present in $Android_{j2k}$.

\textbf{Step 3. Identification of file migration.}
In order to find real cases of migrations, we navigated through all commits from the repositories identified in step 2.
Then, we applied the following procedure: consider that a repository is a set of versions (commits) $C_r = \{c_i, c_{i+1}, ..., c_n\}$ where $i$ determines the commit number, i.e., $c_1$ is the first commit and $c_n$ is the last commit.
Then, to find migrated files, we compared consecutive commits, $c_i, c_{i+1}$ to extract a pair of files, $f_i, f_{i+1}$, which should respect the following conditions:
\begin{inparaenum}[i)]
\item $f_i$ is a Java file from $c_i$ and was removed in $c_{i+1}$,
\item $f_{i+1}$ is a Kotlin file added in $c_{i+1}$, and
\item $f_i$ and $f_{i+1}$ share the same filename ignoring the file extension (.java, .kt).
\end{inparaenum}

As Table~\ref{cap6:tab:datasets} shows, from GitHub projects, we identified \numprint{7275} commits with \numprint{27375} migrated files from \numprint{1179} projects. 
These commits form the $GitHub_{j2k}$ dataset.
From FAMAZOA,  we found \numprint{3118} commits with migrations that migrated \numprint{8754} files from 266 projects. 
These commits form the $Android_{j2k}$ dataset.

%We ended with 270 applications in $Android_{j2k}$ and 1187 projects in $GitHub_{j2k}$.
%Finally, we found \numprint{3118} commits with migration that migrated \numprint{8754} files, as Table~\ref{cap6:tab:datasets} shows.

%We applied steps 2 and 3 presented in Section~\ref{sec:met:data}, and we identified 270 out of 387 (69\%) applications with at least one migration from Java to Kotlin.

%\paragraph{$Android_{j2k}$: dataset of Android applications with migrations}
%\label{sec:dataset:androidj2k}

%To build our dataset of Android applications with migrations, we mined the repositories of FAMAZOA v3~\cite{FAMAZOA}.
%FAMAZOA is a dataset of open-source applications written in Kotlin, and it contains 387 applications written partially or totally in Kotlin collected from 3 dataset of Android open-source applications: AndroidTimeMachine~\cite{Geiger2018:data}, AndroZoo~\cite{Allix2016} and F-Droid.\footnote{F-droid (repository of open-source Android applications):  \url{http://f-droid.org}}
%We applied steps 2 and 3 presented in Section~\ref{sec:met:data}, and we identified 270 out of 387 (69\%) applications with at least one migration from Java to Kotlin.
%Since FAMAZOA includes applications hosted on GitHub, to avoid duplicates, we removed 170 applications from $GitHub_{j2k}$ that are present in $Android_{j2k}$.
%We ended with 270 applications in $Android_{j2k}$ and 1187 projects in $GitHub_{j2k}$.
%Finally, we found \numprint{3118} commits with migration that migrated \numprint{8754} files, as Table~\ref{cap6:tab:datasets} shows.

\begin{table}
    %\footnotesize
    \centering
    \caption{Results of the data extraction.} 
    \begin{tabular}{lccc}
        %\rowcolor{gray!50}
        \toprule
        \multirow{2}{*}{\textbf{Dataset}} & \textbf{\#Projects }&\textbf{\#Commits} & {\textbf{Migrated files in commits }} \\
         
        &  \multicolumn{2}{c}{\textbf{with migrations}}&\\ 
        \midrule
        $GitHub_{j2k}$ &  \numprint{1179} & \numprint{7275}  & \numprint{27375}\\
        $Android_{j2k}$ & 266&  \numprint{3118} & \numprint{8754} \\
        \bottomrule
    \end{tabular}
    \label{cap6:tab:datasets}
\end{table}

\begin{figure}
%Scripts: /Users/matias/develop/kotlinresearch/kotlinmigrationdiff-research/learn-to-migrate/git-Learn-to-migrate/machine-learning/ranking/StatsProjects.py
     \centering
     \begin{subfigure}[b]{0.66\textwidth}
         \centering
         \includegraphics[width=\textwidth]{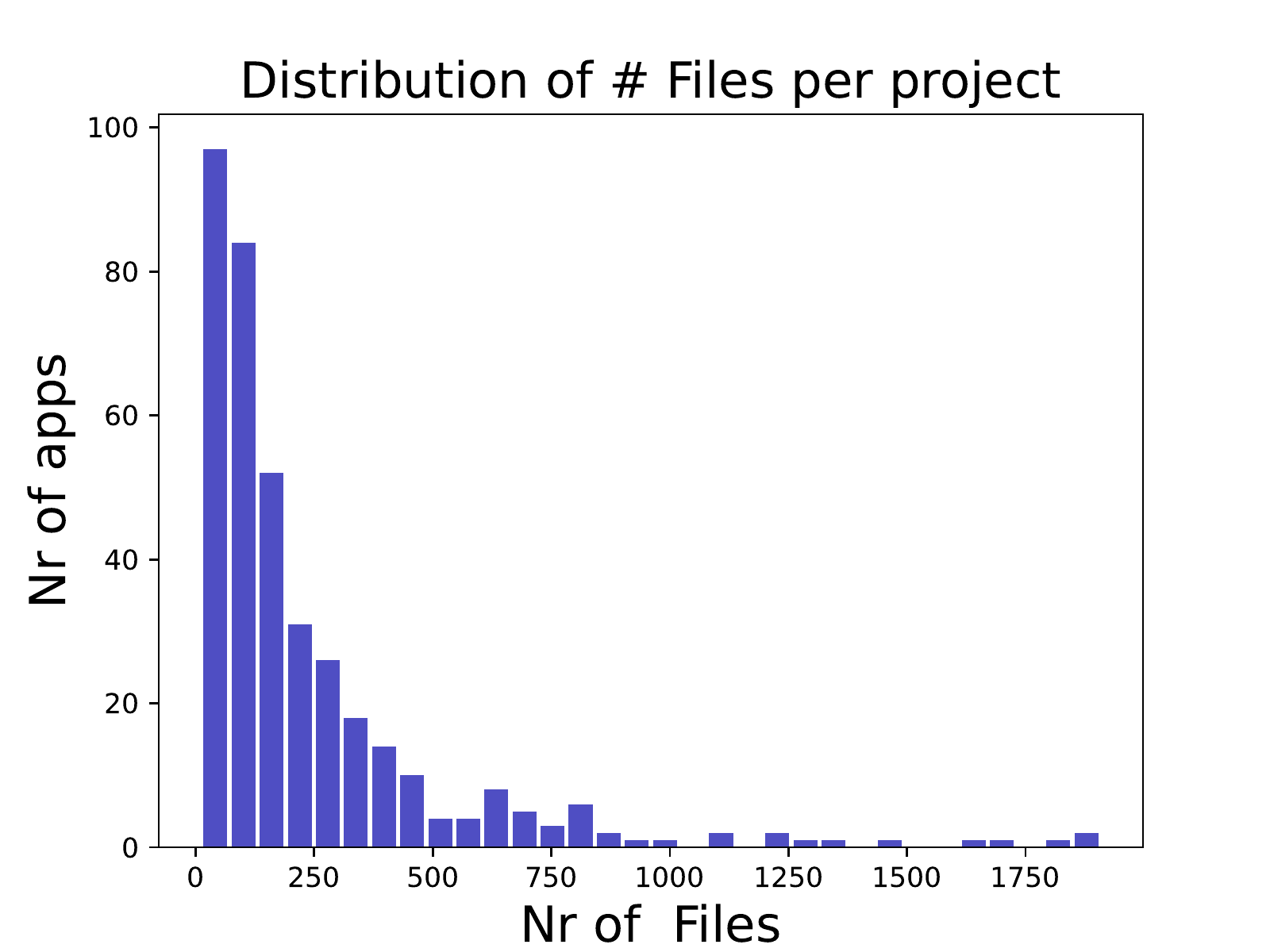}
         \caption{Number of files}
         \label{fig:distNrFilesAndroid}
     \end{subfigure}
     \hfill
     \begin{subfigure}[b]{0.66\textwidth}
         \centering
         \includegraphics[width=\textwidth]{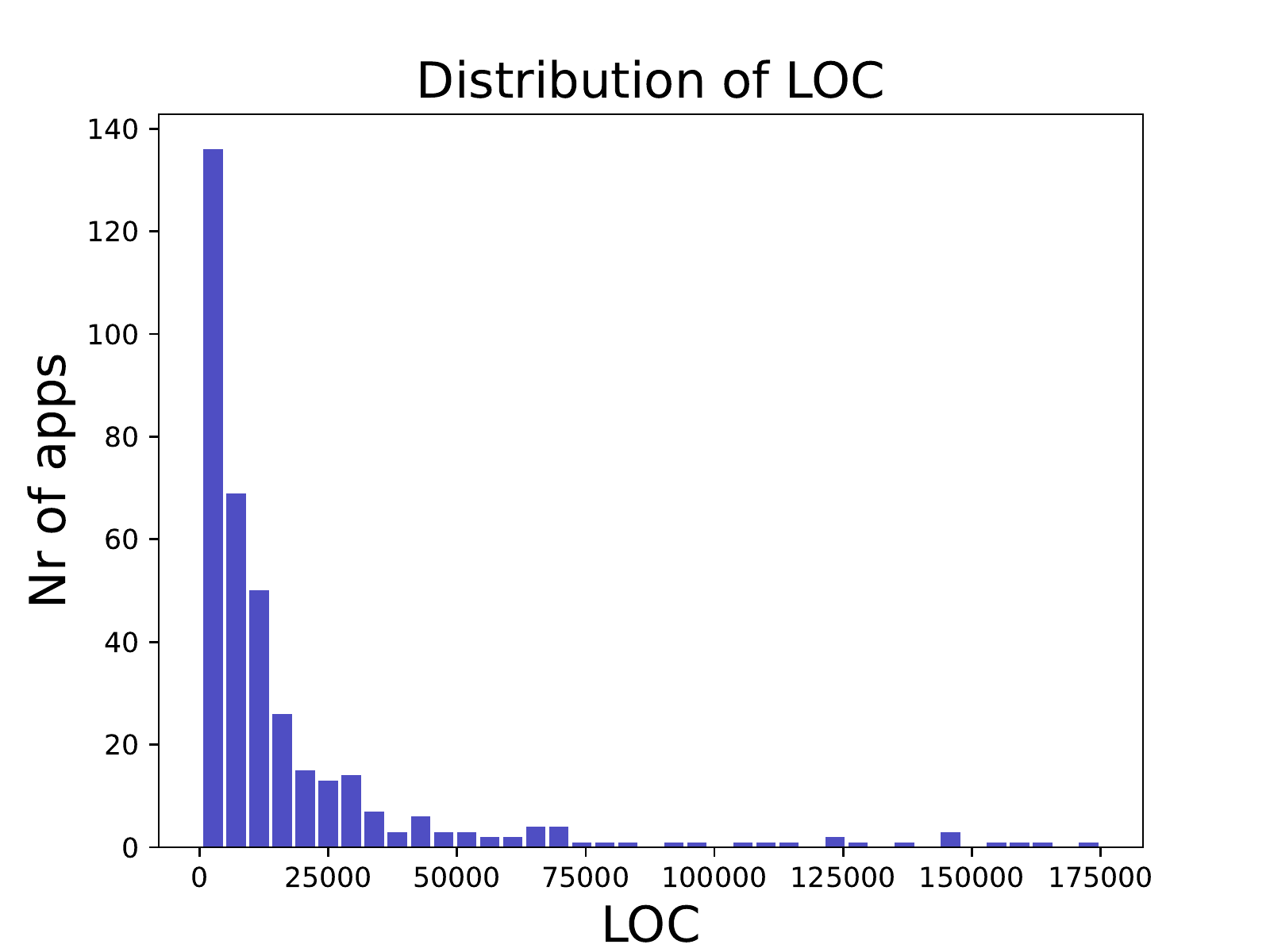}
         \caption{LOC}
         \label{fig:distLOCAndroid}
     \end{subfigure}
     \hfill
     
     \begin{subfigure}[b]{0.66\textwidth}
         \centering
         \includegraphics[width=\textwidth]{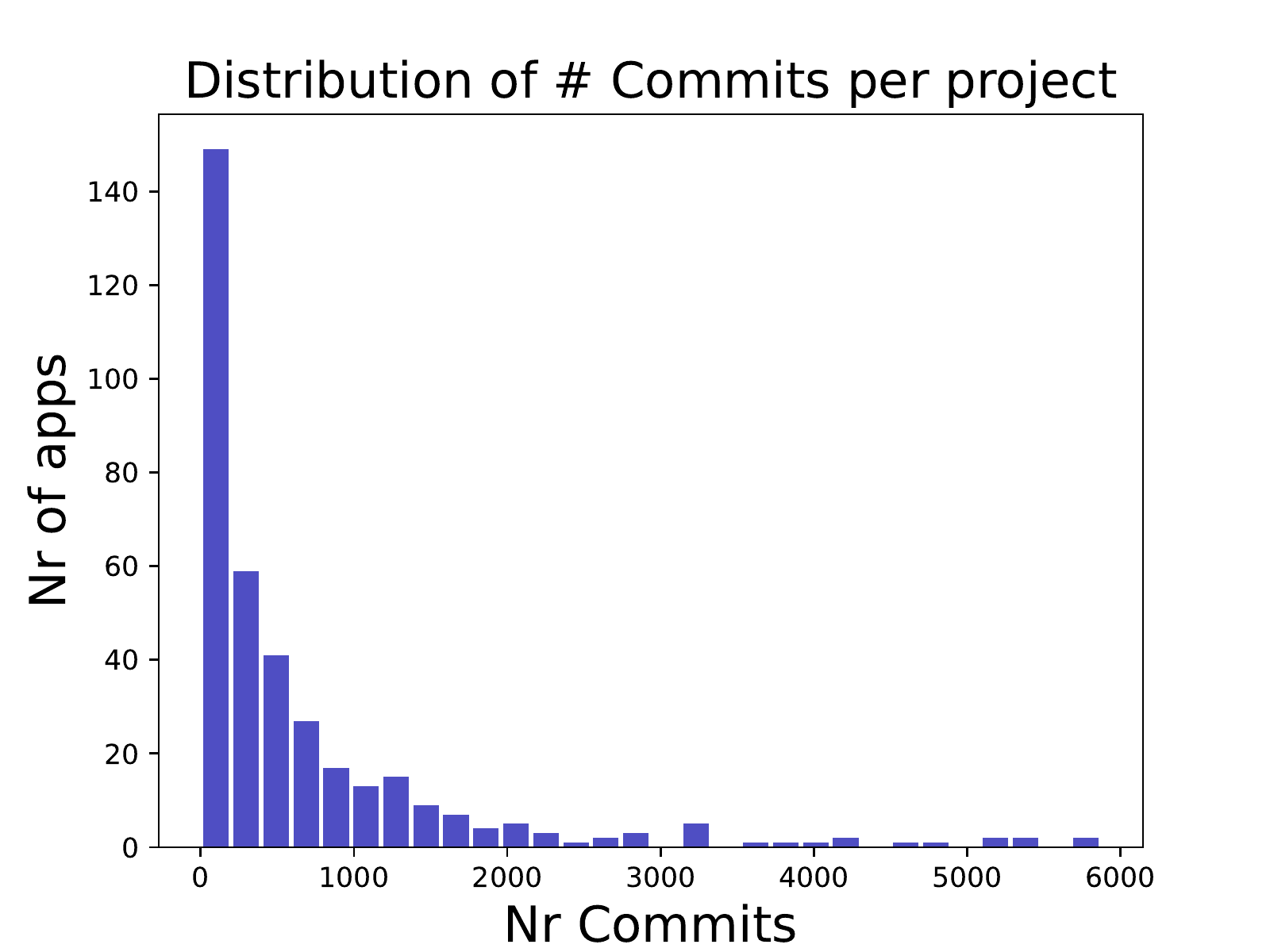}
         \caption{Number of commits}
         \label{fig:distNrCommitsAndroid}
     \end{subfigure}
 
        \caption{Number files, LOC and commits from $Android_{j2k}$'s project.}
        \label{fig:distibutionFilesCommits}
\end{figure}

\paragraph{Statistics about selected projects}

We now discuss some statistics about the selected projects.
We report the results from the $Android_{j2k}$ dataset, as we have observed similar results from the analysis of the $GitHub_{j2k}$ dataset (e.g., similar distributions).

Figure \ref{fig:distNrFilesAndroid} shows the distribution of the number of files. 
It corresponds to a long-tailed distribution: most of the projects have fewer than 500 files (median 139, mean 243). 
The distribution of lines of code (LOC) presented in Figure \ref{fig:distLOCAndroid} also follows the long tail distribution: most of the projects have less than 25k LOC (median 7937, mean 18235), and few projects with a much larger size in terms of LOC.
Finally, Figure \ref{fig:distNrCommitsAndroid} shows the distribution of number of commits: Most of the project has fewer than 1000 commits (median 308, mean 694).

\begin{figure}
%Scripts: /Users/matias/develop/kotlinresearch/kotlinmigrationdiff-research/learn-to-migrate/git-Learn-to-migrate/machine-learning/ranking/StatsProjects.py
   
         \centering
         \includegraphics[width=0.80\textwidth]{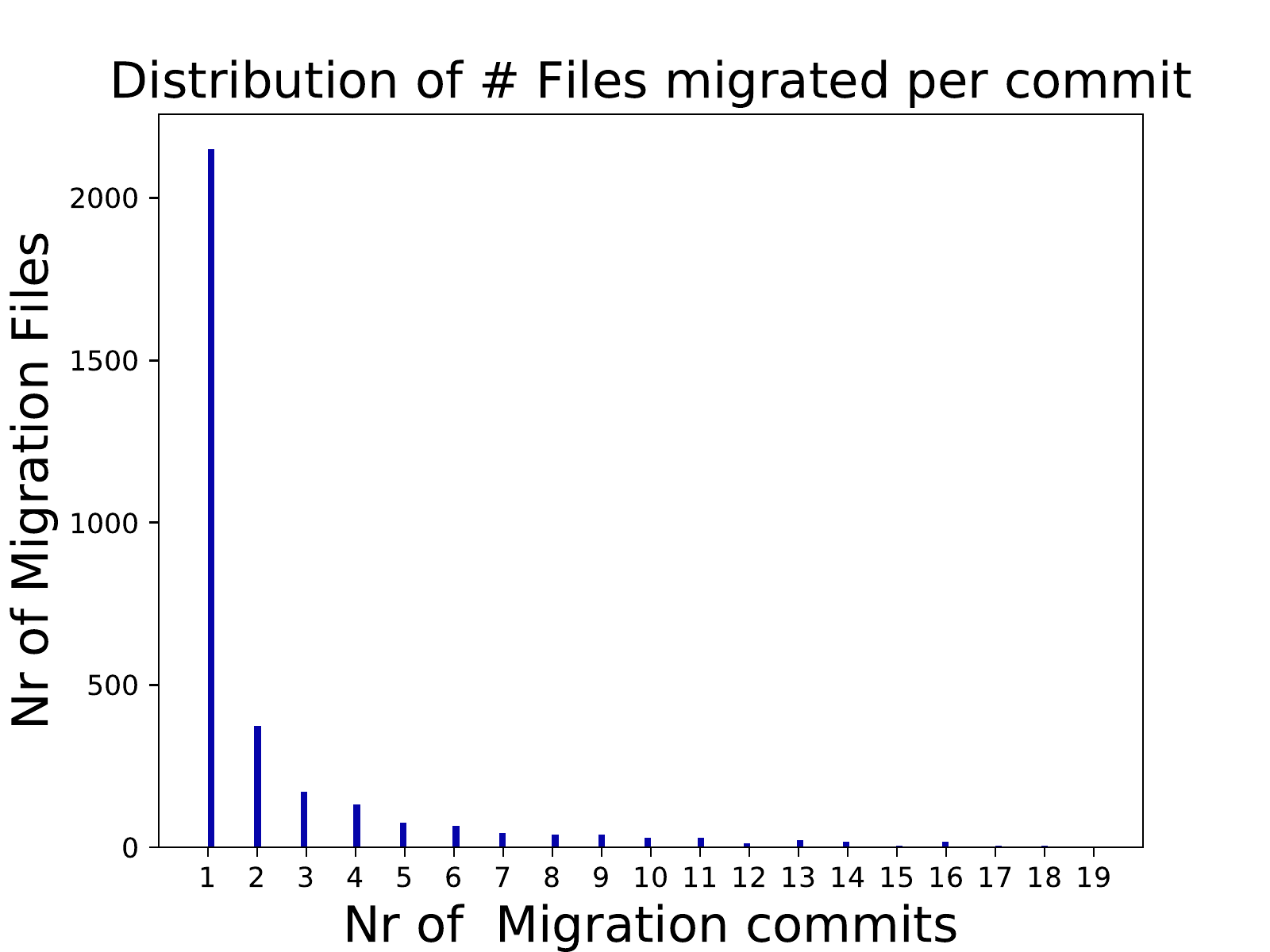}
        \caption{Distribution of number of files migrated by migration commits from $Android_{j2k}$ projects. }
    \label{fig:distMigFilesCommit}
\end{figure}

Regarding the commits, we also inspected the distribution of the migrated files (according to the heuristic presented in Section~\ref{sec:met:data}) over the commits that carry out, at least, one migration.
Figure~\ref{fig:distMigFilesCommit} shows the distribution: we observe that most of the commits migrate exactly one file, or very few files (less than 5).
This shows that developers of the analyzed apps do incremental migrations: they migrate one file and then commit the migrated code.

% /Users/matias/develop/kotlinresearch/kotlinmigrationdiff-research/kotlin-migration-research/executorScriptMigA/checkCLocFiles.py

\subsection{Feature extraction}
\label{sec:features_extraction}

\approachJtK{} relies on {56} metrics extracted from the source code of open-source projects with file migrations from Java to Kotlin.
To extract 12 exclusive Android metrics, we built a static analysis tool using Spoon~\cite{spoon}. 
The remaining 44 source code metrics are extracted using CK~\cite{aniche-ck}, which also applies static analysis to calculate the code metrics.

To extract these metrics from the files of each commit with migration, we created a tool that takes as input
\begin{inparaenum}[\it a)]
\item a Git repository and 
\item the list of commits with migration.
\end{inparaenum}
This tool relies on jGit, a pure Java library that implements the Git version control system.\footnote{jGit: \url{https://www.eclipse.org/jgit/}}
The tool clones the software repository, then navigates through all commits.
Let $C_r = \{c_1, c_2, ..., c_n\}$ be the set of commits with migrations of a given repository.
$\forall c, c \in C$ the tool checks out the source code, then extracts the metrics by calling CK~\cite{aniche-ck} and our Android features detector.
When a repository is analyzed, our tool generates a JSON file which has, for each commit, the values for features extracted grouped by file affected by the commit.

\subsection{\approachJtK{} training}
\label{sec:met:training}

The model used by \approachJtK{} was trained using a learning-to-rank algorithm.
The existing learning-to-rank algorithms are categorized into three approaches: pointwise, pairwise, and listwise~\cite{Liu2009}.
In the pointwise approach, the input is a single document.
Consequently, it does not consider the inter-dependency among documents~\cite{Liu2009}.
On the other hand, pairwise and listwise algorithms consider the inter-dependency among documents.
In the pairwise approach, the ranking problem is reduced to a classification problem on document pairs, whereas the listwise approach addresses the ranking problem by taking ranking lists as instances in both learning and prediction~\cite{li2011short}.

In the context of incremental migration that we target in this paper,
we hypothesize that the decision to migrate or not one file is made considering a project's context and not a file individually.
For instance, in a migration step $S$ given by commit $C$, a developer chooses a set of files $FM$ (one or more) to be migrated over other files $NFM$ that are not migrated in that step.
Thus, to capture that decision between files to migrate, we decide to use the \emph{pairwise} approach. 
During model training, that approach considers that, in the query associated with commit $C$, file $f_i$ from $FM$ was ranked higher than a file $f_j$ from $NFM$.

We trained our model using LambdaMART~\cite{burges2010}, an algorithm developed by Microsoft that applies the pairwise approach and has been shown to be among the best performing learning methods based on evaluations on public datasets~\cite{Ganjisaffar2011}.
We used the LambdaMART implementation provided by XGBoost, a scalable machine learning system for tree boosting proposed by Chen et al.~\cite{Chen2016XGboost}.\footnote{This implementation can also perform listwise ranking. However, as shown in our appendix, its pairwise version outperforms its listwise version.}
Given a query done on XGBoost, this tool outputs as \emph{predicted relevance} values (see Section \ref{cap6:sec:ranking:serving}) a float number per document, where a higher value means higher relevant.
The model was trained with the information extracted from \numprint{7275} commits with at least one file migration from the $GitHub_{j2k}$ dataset.
These commits have \numprint{1495734} files where \numprint{27375} were migrated, as Table~\ref{cap6:tab:datasets} shows.

%%%%
%%% Commented for the moment, it does not give many information
%\begin{figure}[t!]
%    \centering
%    \includegraphics[width=0.50\textwidth]{ranking_evaluation.png}
%    \caption{For each commit that migrates code (e.g., commit \#2), our approach generates a ranking with all project files. This is shown as the \emph{Recommendation list} on the figure.
%    This ranking is evaluated based on the \emph{relevant} documents (i.e., those migrated by the developers).
%}
%    \label{cap6:fig:ranking_evaluation}
%\end{figure}  

\subsection{\approachJtK{} evaluation}
\label{cap6:sec:ranking:evaluation}

This section presents an evaluation of the performance of the trained model.
For that, we used $Android_{j2k}$ (see Section \ref{sec:met:data}) as the testing dataset.

The performance of \approachJtK{} was compared to that one from those baselines, which we present in Section \ref{sec:methodology:baselines} using the metrics presented in Section \ref{sec:methodology:metrics}.

\subsubsection{Baselines}
\label{sec:methodology:baselines}

To our knowledge, there was no baseline on code recommendation migration that we can take to compare our approach.
For this reason, we created two baselines.

First, we defined a \emph{Random} baseline, which is implemented by a recommendation approach that randomly recommends files to migrate.
Although it is not a realistic strategy, we included it in this experiment to prove that our approach surpasses random choices.

Second, we defined a baseline named \emph{\googlestrategy{}} based on Google's guideline for migrating Android apps to Kotlin~\cite{AndroidDevelopers2020a}. 
This guideline suggests migrating first data model classes, then test classes, followed by utility methods, and finally other classes such as fragments and activities.

To create a baseline based on Google's guideline, 
we implemented a recommendation approach with the same interface that \approachJtK{}: given a project $P$, it produces a list of candidate files to migrate.
The implementation generates that list based on the features extracted from the project's files (see Section \ref{cap6:sec:design:features}).
It considers feature \textit{isPOJO} to identify data models and the feature \textit{isTest} to find test files.
Moreover, it classifies a file as a \emph{utility} file when it only has static methods, i.e., the number of methods $> 0$ and the number of methods equal the number of static methods.

\subsubsection{Metrics}

\label{sec:methodology:metrics}

%%%%% performace

To evaluate the performance of \approachJtK{} and the baselines, 
we computed the Mean Average Precision at K ($MAP@k$)~\cite{baeza1999modern}.

We now describe how we compute the $MAP$ metric.
Each \emph{data point} from the testing set corresponds to a commit $C_i$ that migrated some pieces of code $M$ from Java to Kotlin. 
Therefore, in the version produced by the previous commit $C_{i-1}$, $M$ is written in Java and in the version produced by $C_i$ $M$ is written in Kotlin.

We used the $MAP$ metric to evaluate the precision of the recommendations made by \approachJtK{} and the baselines for each data point.
We first created a query $Q$ (explained in Section \ref{sec:representation}) for each data point $D_i$, corresponding to commit $C_i$.
This query represents the project's files from the version \emph{previous} to the migration done by $C_i$.
Then, we queried \approachJtK{} given $Q$ as input, which returns a list of suggested files to migrate.
To measure the precision of the suggestion given $Q$, 
we calculated the metric $AP_Q@K$.
It compares the top-K results from that returned list with
the real migration that a developer has made in commit $C_i$.
The files migrated by $C_i$ are the \emph{relevant} documents.
(We recall that a document represents a file from the project under migration, see Section \ref{cap6:sec:ranking:traning}).

$AP_Q@K$ is given by Eq.~\ref{formula:ap}:

%%https://towardsdatascience.com/breaking-down-mean-average-precision-map-ae462f623a52
\begin{equation}
    \label{formula:ap}
    AP_Q@K = \frac{1}{\text{TR}} * \sum_{j=1}^{k}{{Precision}@j \;  * \; rel@K}
\end{equation}

where TR refers to the total number of relevant documents retrieved,  $Precision@K$ is given in Eq.~\ref{formula:prec} and $rel@K$ is given in Eq.~\ref{formula:rel}. The notation $@k$ means, in this context, the first $k$ elements from a list.

% https://towardsdatascience.com/breaking-down-mean-average-precision-map-ae462f623a52
\begin{equation}
\small
 \label{formula:prec}
      {Precision}@n =  \frac{\text{\# of recommended documents @$n$ that are relevant}}{\text{\# of recommended documents @$n$}}
\end{equation}

%%The relevance function is an indicator function which equals 1 if the document at rank k is relevant and equals to 0 otherwise.
\begin{equation}
 \label{formula:rel}
     rel@K = \begin{cases}
     1 & \text{the document at rank K is relevant}\\
     0 &  \text{otherwise}
     %b & d
     \end{cases}
\end{equation}

Finally, the $MAP@K$ of a recommendation system is the mean of the $AP_i@K$ from all queries done (one query per training point):

\begin{equation}
    \text{MAP@K} = \frac{1}{N} \sum_{i=1}^{N}{AP{_i}@K}
\end{equation}

where N is the number of training points (i.e., number of queries done).
$MAP@K$  ranges from 0 to 1. 
A perfect ranking result in $MAP@k$ equals 1.
In this experiment, since the median number of files migrated per commit is 1, we consider $k$ to be ranging from 1 to 10.

%%%%%%%%%Improvement

We computed our approach's performance improvement by comparing \approachJtK{}'s $MAP@K$ performance with that of a baseline approach $B$ using the formula: 
%$Improvement = \frac{O - B}{O}$, 
\begin{equation}
  Improvement = \frac{\text{MAP@K$_{\approachJtK{}}$} - \text{MAP@K$_B$}}{\text{MAP@K$_B$}}
\end{equation}

In this experiment, we report the improvement of \approachJtK{} w.r.t the best baseline between Random and Google recommendation strategies.

%where $O$ denotes the ranking performance of our approach, $B$ means the ranking performance of a baseline ranking schema.

% 3118/3118 commits with migration
% 8754 migrated files

\section{Results}
\label{sec:evaluation}

%\begin{table*}[ht]
\begin{sidewaystable}
\centering
\caption{Mean Average Precision (MAP) at \textit{K} of a random, \googlestrategy{} and \approachJtK{} and strategy.}
\begin{tabular}{l c c c c c c c c c c c}
\toprule
\multirow{2}{*}{Suggestion strategy} & \multicolumn{9}{c}{Mean Average Precison (MAP) at \textit{K}} \\
\cmidrule{2-12}
& \textit{k}: & 1 & 2 & 3 & 4 & 5 & 6 & 7 & 8 & 9 & 10\\
\midrule
{Random} && 0.188 & 0.238 & 0.251 & 0.262 & 0.268 & 0.271 & 0.274 & 0.276 & 0.278 & 0.278\\
\googlestrategy{} && 0.108 & 0.157 & 0.173 & 0.184 & 0.190 & 0.195 & 0.198 & 0.200 & 0.201 & 0.202 \\
\approachJtK{} && 0.225 & 0.262 & 0.280 & 0.289 & 0.293 & 0.297 & 0.301 & 0.305 & 0.306 & 0.308 \\
\hdashline

Improvement vs Random && 19.7\% &	10.1\% &	11.6\% &	10.3\% &	9.3\% &	9.6\% &	9.9\% &	10.5\% &	10.1\% &	10.8\% \\
Improvement vs \googlestrategy{} &&108.3\% &	66.9\% &	61.8\% &	57.1\% &	54.2\% &	52.3\% &	52.0\% &	52.5\% &	52.2\% &	52.5\% \\

\bottomrule
\end{tabular}
\label{cap6:tab:ranking_result}
%\end{table*}
\end{sidewaystable}

\subsection*{RQ: \rqtwoCapVI}

This section presents the results of the evaluation of a random approach and the \approachJtK{} applied to rank file-level migrations. 
Table~\ref{cap6:tab:ranking_result} summarizes our results.

First, the $MAP$ values go from 0.225 to 0.308.
We recall that a system that perfectly suggests the files to migrate would have a MAP equals (or close) to~1.
Although the results show that there is a large room for improvement, we consider that this result is important to settle the first baseline on Java to Kotlin migration. 
In Section \ref{cap6:sec:discussion} we discuss different directions for improving our results.

Second, our results show that when \textit{k} increases, $MAP$ values also increase.
This makes sense since a greater \textit{k} means that the evaluation considers more files from the suggestion list. Thus, the probability of finding a relevant document (a file that was actually migrated) also increases.

The table shows that \approachJtK{} outperforms both baselines.
\approachJtK{} presents an improvement w.r.t. random of around 10\% $\forall k \in [2, 10]$, and the largest improvement (19.7\%) for \textit{k}=1.
Moreover, \approachJtK{} presents an improvement higher that 50\%, $\forall k \in [2, 10]$, reaching an improvement of 108\% for \textit{k}=1.

The low performance of \googlestrategy{} may suggest that the developers that migrated code in the applications we considered did not apply the suggestion from Google's guideline on migration~\cite{AndroidDevelopers2020a}. 
As we mention in Section \ref{cap6:sec:discussion}, more research is needed on measuring the adoption of Google's guideline and the definition of new guidelines from previous migrations to understand how developers migrate Java to Kotlin code.

\begin{tcolorbox}

\textbf{Response to RQ:}~\textit{\rqtwoCapVI} 

The results show that our learning-to-rank approach \approachJtK{} has mean average precision (MAP) between 0.225 and 0.308, and surpasses the performance from two baselines: Random strategy and strategy based on \googlestrategy{}. 
Nevertheless, these results suggest that there is still room for improvement, as the performance is below the ideal prediction performance (that is, MAP $ \simeq 1$).
\end{tcolorbox}

We now present an example that shows how \approachJtK{} has done the prediction and how we evaluate it.

\paragraph{Example}

\begin{figure}[t]
    \center
    \includegraphics[width=0.8\columnwidth]{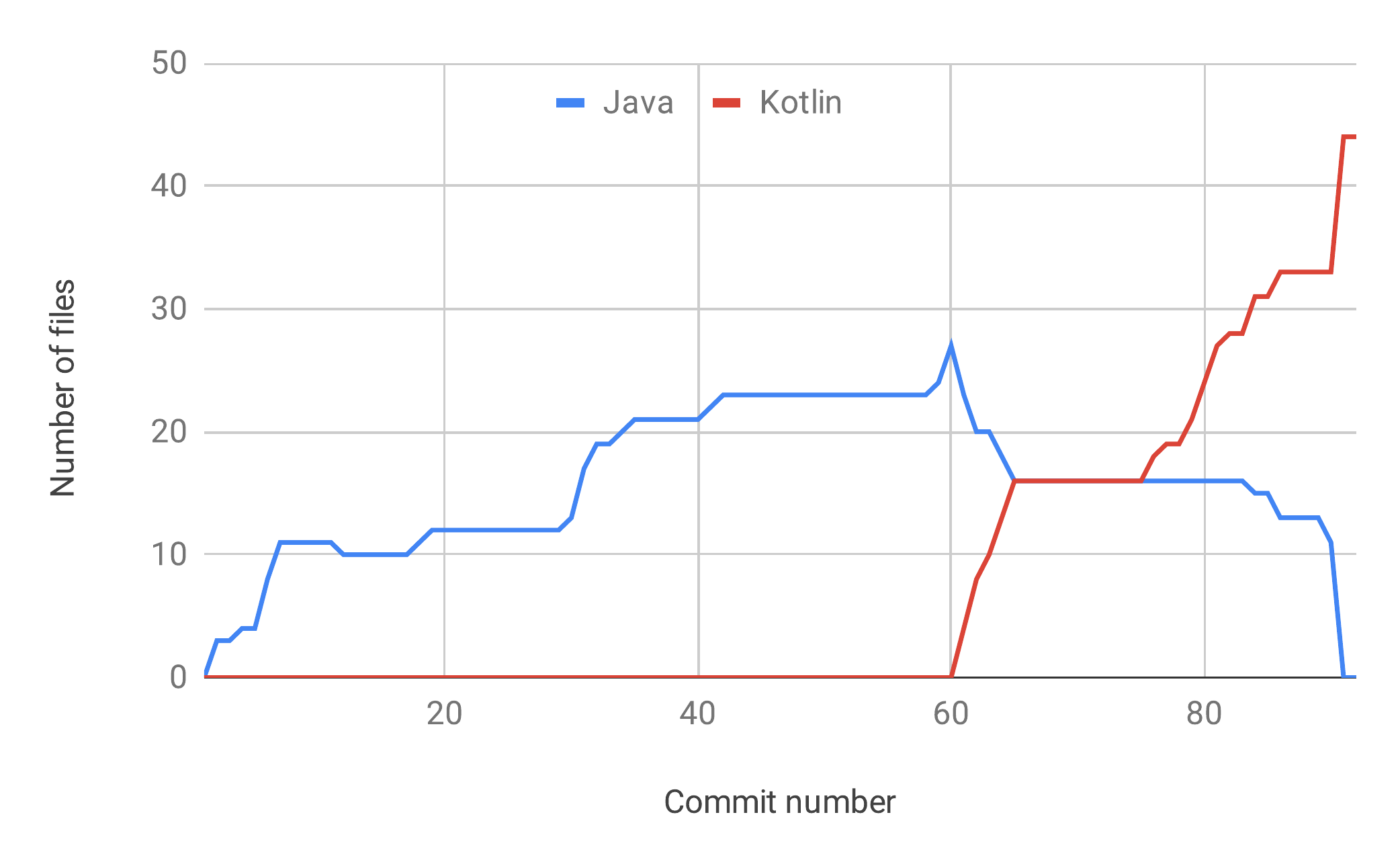}
    \caption{Evolution of the numbers of files written in Java and Kotlin along the incremental migration of Simple Calendar Pro application.
   }
    \label{fig:case1}
\end{figure}

In this example, we focus on the suggestion made by \approachJtK{} during the iterative migration of `Simple Calendar Pro' application\footnote{Simple Calendar App at Google Play store: \url{https://play.google.com/store/apps/details?id=com.simplemobiletools.calendar.pro}} from Java to Kotlin.
Simple Calendar Pro is an application published on Google Store that has more than \numprint{100000} downloads and its source code is hosted on 
GitHub.\footnote{Simple Calendar App at Github: \url{https://github.com/SimpleMobileTools/Simple-Calendar/}}
This application was initially written in Java, but was fully migrated to Kotlin in two months.  
Starting in commit $09ef99$, their developers performed an incremental migration that was completed in commit $eee184$, after 202 commits.
Figure~\ref{fig:case1} shows the number of Java and Kotlin files on each commit from the app along the incremental migration.

We apply \approachJtK{} on a version of Simple Calendar Pro, identified by commit {$2d1c59$}.
In this version, Simple Calendar has {38} Kotlin files (most of them already migrated by previous commits) and {6} Java files, i.e., {6} candidate files to be migrated. 
Table \ref{tab:case_1} presents those Java files.
Given that version of Simple Calendar, our approach generates a \emph{predicted relevance value} (described in Section \ref{cap6:sec:ranking:serving}) for each file. Those are also presented in Table \ref{tab:case_1}. 
Then, it creates a ranked list of these {6} files considering these values.
Therefore, according to \approachJtK{}, {\texttt{AboutActivity.java}} should be the first migrated because it has the highest \emph{predicted relevance value} {(0.96)}, followed by {\texttt{MyWidgetProvider.java} (0.58)}, etc.

Now, we compare this suggestion from \approachJtK{} with the real migration done by the developer on that particular version of Simple Calendar.
The developers migrated only one file, {\texttt{AboutActivity.java}}, and that change produced a new version {(commit $59b020$)} of their application.
In this case, the file in the first position of the list of recommendations made by \approachJtK{} was exactly the same file migrated by the developers, resulting in a MAP@1 equal to 1.

Note that the migration case done by the developer does not follow Google's migration guideline \cite{AndroidDevelopers2020a}, which prioritizes to migrate, for instance, utility classes (\texttt{Utils.java} in this case) over `Activity' files (such as \texttt{AboutActivity.java}).

\begin{table}
\centering
\caption{Comparison between the migration performed by developers (relevant documents are those from migrated files) and the recommendation made by \approachJtK{} (Predicted Ranking), when Simple  Calendar Pro applications at version $2d1c59$ is given as input. 
We recall that for \emph{Predicted relevance} values, higher is better.}
\begin{tabular}{l c c c c}
\toprule
Candidate & Predicted  & Relevant& Predicted \\
files & relevance  & Document & Ranking \\
\midrule
AboutActivity.java &  0.96 & Yes & 1\\
MyWidgetProvider.java & 0.58 & -& 2\\
WidgetConfigureActivity.java & 0.42 & - & 3\\
Utils.java & 0.32 & - & 4\\
LicenseActivity.java & 0.27 & - & 5\\
Constants.java & -0.24 & - & 6\\
\bottomrule
\end{tabular}
\label{tab:case_1}
\end{table}

\begin{figure}[t]
    \center
    \includegraphics[width=0.9\columnwidth]{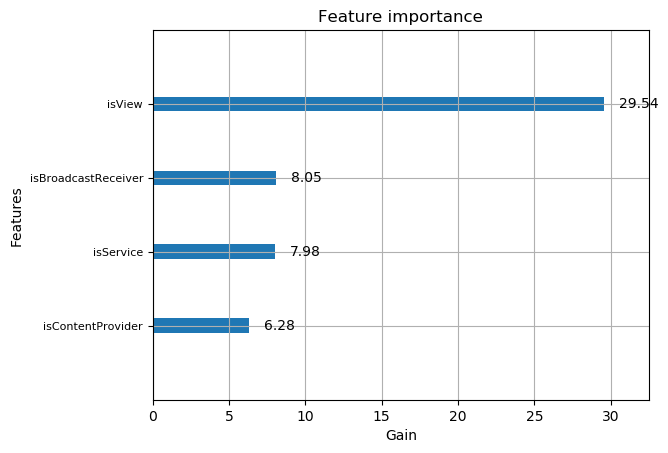}
    \caption{The most important features reported by XGBoost, computed using the 'Gain' importance.
   }
    \label{fig:model_gain}
\end{figure}

\paragraph{Feature Importance}
Figure \ref{fig:model_gain}
shows the importance of the features computed by the XGBoost tool (we recall that it is the tool we use to create the model), as it provides a built-in module to calculate the importance.
The importance provides a score that indicates how useful or valuable each feature was in the construction of the model.
We report the default type of importance: ``gain''.  A higher value of this metric compared to another feature implies that it is more important for generating a prediction.
We observe that the most important feature is $isView$, followed by $isBroadcastReceived$ and $isService$.
This means that our model is mostly influenced by the type of class (isView, isService, isBroadcast) that is under analysis.

\section{Threats to validity}
\label{sec:threatsvalidity}

In this section, we discuss the threats that could affect the validity of our results. 
\subsection{Construct validity}
Threats to construct validity concern the relation between theory and observation.

\paragraph{Quality of  migration suggestion}
In this paper, we present an approach that suggests migration based on a model learned from previous human-written migrations.
However, there is no guarantee that the migrations done by developers used for training and the migrations proposed by our approach correspond to the best migration order.
For determining such cases, it is necessary to first define what is a good migration order, and then to check whether a migration step is then good or not, or there are other better orders.
To our knowledge, no previous work has proposed such a definition or measured the quality of a migration order.

\paragraph{Learning from migrations in open source projects}
To create an accurate machine learning model, a large amount of data is essential. 
Due to the absence of a benchmark dataset of file migration from Java to Kotlin, we mined open source projects from GitHub and FAMAZOA. 
We used this information to train and evaluate our model.
However, there is a risk that open-source projects and not open-source projects might be migrated differently.
Thus, the learned model would not adequately characterize the migration activity of those projects.

\paragraph{Automated evaluation}
To have an automated evaluation process of \approachJtK{}, we consider examples of file-level migrations from open source projects as ground truth.
However, we do not consider the motivation behind these migrations because we cannot automatically retrieve this information from the project repositories. 
%Therefor, we do not know why one file was migrated instead of another.
Consequently, our approach may suggest file-level migrations that do not reflect the decision taken by developers.
Nevertheless, we affirm that this first study aimed to explore whether learning-to-rank can model the problem of recommending file-level migrations.

\paragraph{Feature selection}
The choice of the feature set used to train our learning-to-rank model directly impacts its results, depending on whether these features discriminate adequately, files migrated and non-migrated.
However, to the best of our knowledge, no study establishes a relationship between any metric and source code migration.
For that reason, we target source code metrics used in a wide variety of experiments, such as fault prediction~\cite{Kaur2012,Shatnawi2014}, fault localization~\cite{Sohn2017}, testing~\cite{Eski2011}, defect prediction~\cite{Peng2015}, refactoring prediction~\cite{aniche2020effectiveness} and measuring the quality of object-oriented software~\cite{Singh2013}.
Moreover, we consider 12 exclusive Android features that, according to our experience with Android development, could support the decision to perform a file migration.
Nevertheless, there could exist missing features that better describe the migration activity.

\paragraph{Learning algorithm}

In this paper, LambdaMART was the algorithm chosen to build our ranking model.
However, the choice of the machine learning technique to build a prediction model has a strong impact on performance~\cite{Ghotra2015}.
Thus, using other existing algorithms, our approach could present different performance levels.

\subsection{Internal validity}

Threats to internal validity concern all the factors that could have impacted our results.

%\paragraph{\todo{Learning from "Bad" migrations}}

\paragraph{Learning data}

The data used to train and evaluate \approachJtK{} contains all migrations we found in the datasets of Kotlin applications without applying any filtering.
Our model is learned from all migrations regardless of the strategy adopted by the developers. 
It could be the risk that some developers did not apply the most convenient migration strategy. Thus, the model could be created from a portion of misleading data.
To our knowledge, as we discussed in Section~\ref{cap6:sec:discussion}, no work has focused on migration strategies. Thus, there is not enough knowledge about that in order we could be able to curate the training data.

%\paragraph{In the wild evaluation}
\paragraph{Imbalanced data}

%We did not apply any pre-processing technique in our datasets.

We trained and evaluated our model using highly imbalanced datasets, i.e., there are considerably more instances of the non-migrated files than instances of files migrated.
However, some models may under-perform if trained with imbalanced data~\cite{Hall2012}.

\paragraph{Training parameters}
The choice of parameters for the construction of the model is another threat. 
In this work, we use the default parameters of XGBoost. 
Therefore, for different datasets or metrics, the best parameters might be different, leading to different results.

\paragraph{Types of migrations analyzed}
In this paper, we focus on one type of migration: files are translated from Java to Kotlin one-to-one (that is, for every file in Java there is exactly one file in Kotlin, with the same name).
However, there could exist other migrations that our heuristic is not able to detect, and consequently those are not used for training and evaluating our model.
This involves that our model was trained to make such one-to-one migration suggestions, and not others.
Considering those types of migrations could help improve the generalizability of our approach.

\subsection{External validity}

Threats to external validity concern the generalizability of our findings.

\paragraph{Representativeness of our datasets}
Our work relies on two datasets of open source software. 
However, open source software is a small parcel of the existing software.
This fact may limit the generalization of our findings.

\section{Discussion and future work}
\label{cap6:sec:discussion}

% Summarize result

This work presented a study investigating the feasibility of applying learning-to-rank to build an approach to recommend file-level migrations of Android applications.
The results showed that although our approach overcomes the performance from two baselines: The random strategy and the strategy based on \googlestrategy{}, there is room for improvement.
Nevertheless, we highlight the novelty of our approach and argue that these results establish a baseline for future work.
Moreover, it opens new directions for researchers.
In this section, we list some of them.

\paragraph{On defining strategies of migrations}

In this paper, we used migrations done by developers without filtering them.
In future work, we plan to focus on specific migration strategies.
One of the main challenges is determining what a ``good'' migration strategy would be (expressed, for instance, in terms of ease of migration process, quality of the migrated application, etc.) and what a ``bad'' migration strategy would be. Those strategies could be defined by analyzing metrics from migrated projects or developers' experience.
Once a set of targeted migration strategies is selected, our approach can be trained using data from migrated apps that followed such strategies.
%We do a call for more research on empirical studies of migration strategies.

\paragraph{Migration guidelines}

To our knowledge, there is only one \emph{official} guideline for Android migration, which was created by Google~\cite{AndroidDevelopers2020a}.
It defines the following order of migration:
\begin{inparaenum}[\it 1)]
\item data classes, 
\item test cases,
\item utility classes, and
\item other classes.
\end{inparaenum}
We plan to study whether developers that have migrated code follow or not Google's guideline.
Nevertheless, we consider that our approach could complement this guideline. 
Beyond its usefulness for starting the migration process, 
in our opinion, the guideline is quite general and only prioritizes the types of classes to migrate as described just above.
For example, the guideline indicates to start migrating data classes, but does not specify which one. \approachJtK{} could help developers decide which data class(es) to migrate.
Similarly, it does not specify the order of migration of other types of classes such as fragments, activities and ViewModel.

\paragraph{Interpretability and explainability of recommendations}

One of the main limitations and drawbacks of machine learning techniques, including the learning-to-rank technique, is the lack of transparency behind their behaviors, leaving users with little understanding of how particular decisions are made~\cite{Mengnan2019TechniquesInterpratable}.
To mitigate these problems, researchers have proposed interpretable machine learning techniques, which can generally be grouped into two categories~\cite{Mengnan2019TechniquesInterpratable}:
\begin{inparaenum}[\it 1)]
\item \emph{intrinsic interpretability}: implies constructing self-explanatory models that incorporate interpretability directly to their structures, and \item  \emph{post-hoc interpretability}: requires creating a second model to explain an existing model.
\end{inparaenum}
\approach{} can be complemented with other approaches to create post hoc interpretability. 
For example, SHAP (SHapley Additive exPlanations)~\cite{Lundberg2017SHAP} is an approach to interpret predictions, that is, to explain the output of any machine learning model, including learning-to-rank.
In future work, we plan to integrate \approachJtK{} with the official implementation of the SHAP framework. %, which is compatible with the machine learning framework we use (XGBoost).
Moreover, we plan also to focus on intrinsic interpretable techniques.
For example, Zhuang et al. \cite{Zhuang2021InterpretableRanking} propose an interpretable Learning-to-Rank technique, which could be used by \approachJtK{} instead of the traditional learning-to-rank technique.

%%https://github.com/slundberg/shap

% Hyperparameter tuning
%\todo{if we need space we can remove this as is said in TtoVal}
\paragraph{Hyperparameter tuning}
One strategy to potentially improve our results is to perform a hyperparameter tuning.
Each algorithm has a set of parameters, each having its domain, which may have different types (i.e., continuous, discrete, Boolean and nominal), making the entire set of parameters of an algorithm a large space to explore. 
Consequently, the search for the machine learning algorithm's best parameters is demanding in computation complexity, time, and effort~\cite{ArcelliFontana2016}. 
In future work, we plan to explore different techniques of hyperparameter tuning.

% Data balancing
\paragraph{Data balancing}
Another aspect that researchers can focus on are pre-processing techniques to handle the imbalance of our migration dataset, as they can be more important than the choice of the classifier~\cite{Agrawal2018}.
Despite many real-world machine learning applications, learning from imbalanced data is still not trivial~\cite{PECORELLI2020}.
However, other software engineering studies (e.g, ~\cite{Wang2013Smote}) have used Synthetic Minority Oversampling TEchnique (SMOTE) to fix the data imbalance.
%,Tan2015,Bennin2018,Catolino2019
As feature work, we intend to explore pre-processing techniques to understand how they impact the recommendation of file-level migrations.

% Feature engineering 

\paragraph{Feature engineering}
Since our machine learning models achieve a modest performance, we intend to focus on feature engineering as future work.
Adding new features or discarding existing ones could result in a better set of features that may improve our results.
Therefore, more research should be conducted to \begin{inparaenum}[i)]
\item evaluate the current set of features and possibly discard some feature,
\item verify to what extent existing metrics applied in other domains of software engineering, like process metrics~\cite{Yang2015,Hoang2019}, code smells~\cite{Catolino2020} and ownership metrics~\cite{Bird2011,aniche2020effectiveness}, are suitable for our problem and
\item develop new metrics able to characterize better migrated or non-migrated file instances.
For example, new features can be added that reflect the coupling between already migrated and not yet migrated files.
\end{inparaenum}

\paragraph{Recommending groups of files}

A future direction we aim to explore is the suggestion of migration of groups of files: there each suggestion item in the recommendation list would correspond to a set of files that should be migrated together.
There could exist different criteria for grouping such files: such as coupled files, classes and their corresponding test cases, etc.

\paragraph{Granularity of recommendations}

In this paper, we focus on recommendations of \emph{files}.
Nevertheless, our approach could also be adapted for doing suggestions at a different granularity, e.g., \emph{packages}.

\paragraph{Types of migration recommendations}
The proposed approach was trained from one type of migration (one-to-one file, as explained in Section \ref{sec:met:data}), therefore, the recommendations are based on this type of migration.
If other types of migrations are defined, our approach could be extended in order to suggest, in addition to the files that can be migrated in one migration step, the type of migration for each file.
For that, it would be necessary to first define such new types of migrations, and then to mine samples of those migrations from migrated applications in order to train the model.

\paragraph{Feedback from developers}
In this paper, we use a ranking metric ($MAP$) to automatically assess the quality of the recommendations generated by \approachJtK{}. To complement our evaluation, as future work, we plan to conduct a study in which developers who want to migrate their applications would evaluate the recommendations made by our approach.

\paragraph{Deploying \approachJtK{}~in the wild}
We aim to make \approachJtK{} a production-ready model to integrate it with Android Studio, the official IDE for Android development.
To this end, we intend to develop a plugin for Android Studio and make it publicly available in the official JetBrains Plugin Repository,  as Iannone et al.~\cite{Iannone2020} have done.
We believe that by making our approach publicly available, we can receive feedback from users to improve it.

\section{Related work}
\label{sec:relatedwork}

\paragraph{The adoption of Kotlin}

Oliveira et al.~\cite{Oliveira2020} performed a study to understand how developers deal with the adoption of Kotlin in Android development, their perception of the advantages and disadvantages related to its usage.
They found that developers believe that Kotlin can improve code quality, readability, and productivity.
Gois Mateus and Martinez~\cite{GoisMateus2019} have found that 11\% of the Android open-source applications studied have adopted Kotlin.
As a difference from them, our work focuses on a deeper aspect of the adoption of Kotlin, the \emph{migration} of Android applications from Java to Kotlin.

%\paragraph{Empirical studies on Kotlin code}

Researchers have recently conducted different studies on the use of Kotlin (e.g., \cite{Flauzino2018,GoisMateus2020, ARDITO2020106374}). 
Their results present some benefits of adopting Kotlin. For instance, this produces shorter programs and code with fewer code smells than Java programs.
%Flauzino et al. \cite{Flauzino2018} have studied 100 software repositories (not only Android apps) containing Java or Kotlin code (but not both). 
%They found that, on average, Kotlin programs have fewer code smells than Java programs.
%Gois Mateus and Martinez \cite{GoisMateus2020} have studied the adoption of the features introduced by Kotlin. 
%They found that some Kotlin features are more used than others.
%Ardito et al. \cite{ARDITO2020106374} conducted a study with undergraduate students to assess the assumed advantages of Kotlin concerning Java in the context of Android development and maintenance.
%The authors found evidence that the adoption of Kotlin led to a more compact code. 
%Bryksin et al.~\cite{Bryksin:2018:DAK} investigated code anomalies in Kotlin and whether these anomalies could improve the Kotlin compiler.

\paragraph{Migration of Android applications to Kotlin}

Coppola et al.~\cite{Coppola_2019} evaluated the transition of Android applications to Kotlin to understand whether the adoption of Kotlin impacts the success of an application (i.e., popularity and reputation) of Android applications in the App Store. 
Martinez and Gois Mateus~\cite{martinez2021Why} conducted a survey to know why Android developers have migrated Java code to Kotlin. The use of the new features (typically included in modern programming language), not previously fully available using Java,  was one of the most frequently mentioned reasons.
%to identify the main difficulties they have faced. 
Peters et al.~ \cite{Peters2021HowImpact} empirically assessed the impact of migration from Java to Kotlin on the efficiency of the runtime of Android applications.  They found that migrating
to Kotlin has a statistically significant impact on CPU usage,
memory usage, and render duration of frames.
Our work aims to help those developers to do the migration.

\paragraph{Programming language migration}
Previous work has presented approaches to migrate code e.g.,
\cite{Martin2002} C to Java, Cobol to Web \cite{colosimo2009evaluating}, Cobol to Java \cite{Mossienko2003}, C code to Eiffel ~\cite{Trudel2012}.
Other works focus on automated API migrations (e.g., ~\cite{Zhong2010, Nguyen2014, Gu2017}).
Although these works target programming language migrations, none of them focus on migration from Java to Kotlin.

\paragraph{Learning-to-rank applied to software engineering}
Previous work has applied learning-to-rank to software engineering tasks.
For example, on fault localization (\cite{Xuan2014, Le2016, Sohn2017,Kim2019}), bug-finding process (\cite{Ye2014, Zhao2015, Tian2016}), code search (\cite{Niu2017}), defects prediction (\cite{Wang2018,Yang2015b}), rule-specification mining (\cite{Cao2018}), recommendation system to classify and select design patterns \cite{Hussain2019}, third-party libraries~\cite{ALRUBAYE2020106140}. 
Differently from these works, our work is the first to apply learning-to-rank to suggest file-level migrations.

\section{Conclusion}
\label{sec:conclusion}

In this work, we presented \approach{}, an approach to support developers in incremental migration of applications based on supervised machine learning, in particular, the learning-to-rank approach. 
\approach{} produces recommendations for candidate files to migrate and is based on a model learned from real migrations performed by developers.
Despite being a language-independent approach, we evaluate its feasibility in the context of the migration of Android applications from Java to Kotlin.
The results of the evaluation of \approachJtK{} show that {on the task of suggesting files to migrate}, our approach outperforms two strategies considered as baselines.

We believe that our approach may significantly impact Android applications' development because most Android applications are written in Java, and, at the same time, Google is encouraging developers to adopt Kotlin to keep updated their apps with new Android platform features.
%Consequently, \approachJtK{}  is a novel approach that aim to help developers ith the transition from Java to Kotlin.
%we consider that \approachJtK{} can help developers with the transition from Java to Kotlin.

In future work, we plan to integrate \approachJtK{} into a new infrastructure that will aim to help developers along the different stages of migrations such as migration suggestions, code translation, and testing of the migrated code.

\bibliography{references}

\end{document}